\documentclass[appendixfloats]{emulateapj}
\pdfoutput=1
\slugcomment{{\sc Accepted to ApJ:} April 13, 2012} 
\def\Bes{\rm Besan$\c{c}$\rm on}

\usepackage{natbib}
\usepackage{graphicx}

\shorttitle{Maser Extinction Distance Comparison}
\shortauthors{Foster, Stead, Benjamin, Hoare \& Jackson}
\begin{document}
\title{Distances to Dark Clouds: Comparing Extinction Distances to Maser Parallax Distances}

\author{Jonathan B. Foster\altaffilmark{1}, Joseph J. Stead\altaffilmark{2}, Robert A. Benjamin\altaffilmark{3}, Melvin G. Hoare\altaffilmark{2}, James M. Jackson\altaffilmark{1}}


\altaffiltext{1}{Institute for Astrophysical Research, Boston University, Boston, MA 02215, USA \email{jbfoster@bu.edu}}
\altaffiltext{2}{School of Physics and Astronomy, University of Leeds, Leeds, LS2 9JT, UK}
\altaffiltext{3}{University of Wisconsin Whitewater, Whitewater, WI 53190, USA}


\begin{abstract}
We test two different methods of using near-infrared extinction to estimate distances to dark clouds in the first quadrant of the Galaxy using large near infrared (2MASS and UKIDSS) surveys. VLBI parallax measurements of masers around massive young stars provide the most direct and bias-free measurement of the distance to these dark clouds. We compare the extinction distance estimates to these maser parallax distances. We also compare these distances to kinematic distances, including recent re-calibrations of the Galactic rotation curve. The extinction distance methods agree with the maser parallax distances (within the errors) between 66\% and 100\% of the time (depending on method and input survey) and between 85\% and 100\% of the time outside of the crowded Galactic center. Although the sample size is small, extinction distance methods reproduce maser parallax distances better than kinematic distances; furthermore, extinction distance methods do not suffer from the kinematic distance ambiguity. This validation gives us confidence that these extinction methods may be extended to additional dark clouds where maser parallaxes are not available. 
\end{abstract}


\section{Introduction}
The Sun's location in the disk of the Milky Way Galaxy provides, in principle, an excellent vantage point to study the inner workings of a spiral galaxy. Stellar populations can be resolved, chemical abundances and radial velocities can be obtained on a star-by-star basis, and the vertical, radial and azimuthal distribution of both stars and different phases of interstellar gas can be estimated. One of the principal complications, however, lies in the determination of distances of individual stars, star-forming regions and gas clouds. The possibility of substantial systematic errors in distance determination methods have made ``maps'' of the Galactic stellar and interstellar distributions, particularly the maps of spiral structure, potentially unreliable. 

Complicating this process is the fact that different components of the Galaxy might reasonably be assumed to have different, albeit related, morphologies, as is observed in other spiral galaxies.  \citet{BinneyAndTremaine}, for example, distinguish between {\it bright-star} (star-forming) arms, {\it dust} arms, {\it mass} (stellar density) arms, and (gravitational) {\it potential} arms, each of which may have a different distribution.  (To these four tracers, one might add {\it neutral gas} arms, as concentrations of neutral hydrogen gas can have a distribution that is distinct from the tracers mentioned above.) 

In this paper, we concentrate on the distance to (infrared) dark clouds, dense clouds of molecular gas and dust that are frequently associated with the early stages of star formation. Until recently, only kinematic distances were available for these objects, unless they could be clearly associated with young stellar clusters in which case photometric distances were also potentially available. The past decade has seen the development of new techniques to estimate the distances to dark clouds using the extinction of stellar sources along the line of sight \citep[e.g.][]{Marshall:2009, Kainulainen:2011}. There are many different possible ways of estimating extinction distances, with different underlying assumptions and limitations, but in general these extinction distances hold the promise of producing three dimensional maps of dust distributions independent of kinematics.

A key point to note is that kinematic, photometric, and extinction distances are not absolutely calibrated. Although the systematic application of one method presumably yields useful information about relative distances, the lack of certainty about the accuracy of individual distances has led to some justifiable skepticism about maps derived from these techniques. For example, \citet{Marshall:2009} found good agreement between kinematic and extinction distances in the first quadrant, but a significant (1.5 kpc) systematic difference between these two methods in the fourth quadrant.

The on-going campaign to obtain radio trigonometric parallaxes to a set of masers across the Galactic disk holds the promise to test, and perhaps even to calibrate, other methods for estimating distance. Although the $\sim 200$ parallaxes expected will not be enough to completely map the Galaxy, they {\it may} be enough to calibrate these other methods, or at the very least assess their reliability. \citet{Reid:2009}, for example, have already developed a revised kinematic distance model based on early results. If the masers located in the coherent gas structures found in the longitude-velocity diagrams, \citep[e.g.][]{Dame:2001}, show systematic offsets between parallax and kinematic distances, it may be possible to reliably map the entire gaseous structure. This would be similar to how photometric distances---and more recently, parallax distances---of Perseus Arm star formation regions have been used to estimate the kinematic distance correction necessary to correctly map this structure \citep[e.g.][]{FosterMacWilliams:2006}. It is important to note, though, that the intrinsic velocity dispersion of clouds will always produce random errors in the kinematic distances that depend on longitude and radial velocity. Moreover, the near-far kinematic distance ambiguity will remain, and even corrected kinematic distance will be unreliable for sources close to $l=0^{\circ}$ and $l=180^{\circ}$. 

Extinction distances have a different set of selection biases and systematic errors from those associated with kinematic distances. In this work, we use maser parallaxes to test extinction distances. This is possible because masers are generally associated with star-forming regions. Methanol masers, in particular, are associated with high-mass star formation in dark clouds. In this work, we present a pilot study, examining the distance predictions for eleven clouds between $l=5-60^{\circ}$ for which we have near- and mid-infrared data, molecular line emission, and sub-millimeter observations. In \S\ref{SecParallax}, we describe our sample of clouds and their parallax distance measurements. The kinematic distances to these clouds (and their uncertainties) are discussed in \S\ref{SecKinematic}. In \S\ref{SecBlueNumber} and \S\ref{SecRedGiant} we present two extinction distance techniques which we compare to the maser parallax distances. The results of the comparison are outlined in \S\ref{SecComparison}.  

\section{Maser Parallaxes}
\label{SecParallax}
Trigonometric parallax distances are largely free of astrophysical assumptions and therefore constitute an ideal source of information about the structure of our galaxy. Recently, several groups have used Very-Long-Baseline Interferometry (VLBI) phase-referencing techniques to measure the parallax to masers in regions of massive star-formation with an accuracy of better than 10 $\mu$as. In the sample of maser parallax distance examined in this study the uncertainty on the distance is less than 25\% out to 6 kpc. The groups involved include the Bar and Spiral Structure Legacy Survey \citep[BeSSeL;][]{Brunthaler:2011} using the VLBA (Very Long Baseline Array), a survey using the European VLBI Network (EVN) \citep{Rygl:2010}, and the VLBI Exploration of Radio Astronomy (VERA) project \citep[e.g.][]{Choi:2008, Niinuma:2011}. 

These distances have been used to refine Galactic parameters and rotation curves. In particular, \citet{Reid:2009} used maser parallax distances to derive an updated rotation curve and found that the regions hosting massive star formation are systematically rotating 15 km s$^{-1}$ slower than the rest of the galaxy. This result assumes that the determination of the solar peculiar motion (motion with respect to the local standard of rest (LSR)) is well known from \citet{Dehnen:1998}. In the wake of this paper, several studies have attempted new estimates of the solar motion, without convergence to a single value \citep[see comparison tables in][]{Francis:2009, Bochanski:2011, Coskunoglu:2011}. However, some results \citep{McMillan:2010} suggest that the solar peculiar motion in the V direction (V$_{\sun}$) was under-estimated by $\sim$ 5 km s$^{-1}$. This change would reduce the size of the lag in high-mass star-forming regions identified by \citet{Reid:2009} to 10 km s$^{-1}$, but not reduce it entirely.

For this study, we select a subset of the dark clouds with maser parallax distances. Our methods for estimating extinction distances rely upon accurately establishing the boundaries of the densest portion of the cloud. Because a wealth of data exists in the 1st quadrant of the Galaxy, we limit ourselves to masers in this region. Our ideal candidates are clouds with UKIDSS (UKIRT Infrared Deep Sky Survey) Galactic Plane Survey \citep[GPS;][]{Lucas:2008}, Bolocam Galactic Plane Survey \citep[BGPS;][]{Aguirre:2011}, and Galactic Ring Survey \citep[GRS;][]{Jackson:2006} data. BGPS data can be used to identify the boundaries of a cloud; GRS data can be used to assign a velocity, and to check that a BGPS-identified cloud is coherent in velocity; UKIDSS-GPS data offers higher precision near-infrared photometry than the Two Micron All Sky Survey \citep[2MASS;][]{Skrutskie:2006}. Table~\ref{masers} and Table~\ref{clouds} list all the clouds and masers we consider in this study, although some methods can only be applied to a subset of these clouds. Additional maser parallax distances are continually being reported in the literature \citep[e.g.][]{Nagayama:2011, Sanna:2012, Xu:2011} and could provide extra sources for testing distance determinations.

\begin{deluxetable*}{lcccll}
\tablecaption{Masers Used in this Study}
	\tabletypesize{\footnotesize}
	\tablehead{
	\colhead{Name} & \colhead{l} & \colhead{b} &
	\colhead{Distance [kpc]} 
	& \colhead{Reference} & \colhead{Type}}
	\startdata
	G5.89$-$0.39	&	5.8842 	&	-0.3924 	&	$1.28_{-0.08}^{+0.09}$		&	\citet{Motogi:2011}	&	22 GHz H$_2$O \\
	G9.62$+$0.20	&	9.6211 	& 	0.1959 	&       $5.2_{-0.6}^{+0.6}$			&	\citet{Sanna:2009}	&	12 GHz methanol\\
	G23.01$-$0.41 &	23.0096	&	-0.4105	&	$4.59_{-0.33}^{+0.38}$		&	\citet{Brunthaler:2009} &	12 GHz methanol\\
	G23.44$-$0.18	&	23.4398	&	-0.1822	&	$5.88_{-0.93}^{+1.37}$		&	\citet{Brunthaler:2009} &	12 GHz methanol\\
	G23.66$-$0.13	&	23.6566	&	-0.1272	&	$3.19_{-0.35}^{+0.46}$		&	\citet{Bartkiewicz:2008} &	6.7 GHz methanol\\
	G34.39$+$0.22 &	34.3940 	&	0.2215	&	$1.56_{-0.11}^{+0.12}$		&	\citet{Kurayama:2011}	& 22 GHz H$_2$O\\
	G35.20$-$0.74	&	35.1970	&	-0.7431	&	$2.19_{-0.20}^{+0.24}$		&	\citet{Zhang:2009b}		& 12 GHz methanol\\
	G35.20$-$1.74	&	35.2002	&	-1.7364	&	$3.27_{-0.42}^{+0.56}$		&	\citet{Zhang:2009b}		& 12 GHz methanol\\
	W51 Main/S	&	49.4884	&	-0.3879	&	$5.41_{-0.20}^{+0.21}$		&	\citet{Sato:2010}		& 22 GHz H$_2$O \\
	IRAS 19213$+$1723 & 52.1005 &  1.0429	&	$3.98_{-0.50}^{+0.67}$		&	\citet{Oh:2010}			& 22 GHz H$_2$O \\
	G59.7$+$0.1	&	59.7828	& 	0.0647	&	$2.16_{-0.09}^{+0.10}$		&	\citet{Xu:2009}			& 12 GHz methanol\\
	\enddata
	\label{masers}
\end{deluxetable*}

\begin{deluxetable*}{lllllcc}
\tablecaption{Clouds Containing Masers Used in this Study}
	\tabletypesize{\scriptsize}
	\tablehead{
	& & \multicolumn{3}{c}{Cloud Velocity} & & \\
	\colhead{Name} 
	& \colhead{Description\tablenotemark{a}}
	& \colhead{V$_{LSR}$ [km s$^{-1}$]} & \colhead{Reference} & \colhead{Tracer}
	& \colhead{GRS}
	& \colhead{BGPS}
	}
		\startdata
	G05.89$-$0.39	&	UCHII	& $+$7 &	\citet{Arikawa:1999} & CO		& N	&	Y\\
	G09.62$+$0.20	&	HCHII	& $+$4 &	\citet{Scoville:1987} &	 CO	& N	&	Y	 \\ 
	G23.01$-$0.41	&	HMSFR	& $+$77 & \citet{Codella:1997}	& NH$_3$	&	Y	& Y \\
	G23.44$-$0.18 &	HMSFR    & $+$99 & \citet{Codella:2010} & NH$_3$	&	Y 	& Y \\
	G23.66$-$0.13	&	HII region?  & $+$80 & \citet{Urquhart:2011} & NH$_3$ & Y & Y \\
	G34.39$+$0.22 &     HMSFR   & $+$57 & \citet{Bronfman:1996} & CS &	Y 	&	Y\\
	G35.20$-$0.74	&	HMSFR	& $+$35	& \citet{Solomon:1987} & CO & Y & N \\
	G35.20$-$1.74	&	UCHII	& $+$43	& \citet{Stutzki:1984}		& NH$_3$	&	N	&	N \\
	W51 Main/S	&	HMSFR	& $+$55  	& \citet{Zhang:1998}		& CS	&	Y 	& 	Y	 \\
	IRAS 19213$+$1723 & HII region?	& $+$42	& \citet{Bronfman:1996}	& CS 	&	Y 	& 	N	\\
	G59.78$+$0.06	&	HMSFR	& $+$29		& \citet{Bronfman:1996}	& CS		&	N 	&	Y	\\
	\enddata
	\tablenotetext{a}{UCHII -- Ultra-Compact HII Region, HCHII -- Hyper-Compact HII Region, HMSFR -- High-Mass Star-Forming Region.}
	 \label{clouds}
\end{deluxetable*}

\section{Kinematic Distances}
\label{SecKinematic}
Kinematic distances, estimated by assuming gas is in circular rotation, are simple to calculate, but have several sources of systematic error.  Four difficulties with this method are (1) obtaining an accurate rotation curve, (2) correcting for solar motion, (3) breaking the ``near-far'' distance degeneracy in the inner galaxy (where one velocity can map to two distances), and (4) allowing for non-circular motions in the vicinity of well-organized galactic structures. In addition, the uncertainties associated with kinematic distances depend strongly on longitude. 

In order to capture some of the systematic uncertainties associated with choice of rotation curve and solar motion, we calculate kinematic distances to the maser clouds using two different Galactic rotation curves. Our first set of kinematic distances use the \citet{Clemens:1985} rotation curve, which includes a correction to V$_{LSR}$  of the form ($7\sin{l}$) km s$^{-1}$ (which is in the same direction as the recent increase in V$_{\sun}$ of 5 km s$^{-1}$). Our second set of kinematic distances uses the rotation curve proposed by \citet{Reid:2009}, including the 15 km s$^{-1}$ lag for high-mass star-forming regions (effectively a decrease of 15 km s$^{-1}$ in $\Theta_{0}$), since our study uses high-mass star-forming regions. The velocity used for these kinematic distances come from a variety of dense gas tracers in a variety of studies; see Table~\ref{clouds} for the velocities we have used.

We estimate the random error on kinematic distances due to non-circular motion by perturbing the reference velocity by $\pm$ 3 km s$^{-1}$ and deriving the kinematic distance for the perturbed velocities. A velocity spread of 3 km s$^{-1}$ was chosen because this is the azimuthal (i.e. 1D) velocity dispersion of molecular clouds as reported in \citet{Clemens:1985}. Because the far kinematic distance is much larger than the maser parallax distance for all these clouds, we always adopt the near kinematic distance. We note that in the absence of this prior information, HI self absorption can sometimes be used to break the distance ambiguity \citep[e.g.][]{Roman-Duval:2009}.

\section{Blue Number Count Extinction Distances}
\label{SecBlueNumber}
Our first extinction distance method, the Blue Number Count method, uses the number of foreground stars and a Galactic model of the stellar distribution to estimate the distance to dark clouds. Using star counts to estimate the distance to dark clouds goes back at least as far as \citet{Wolf:1923} and \citet{Bok:1931} \citep[see][for a history of the origins of this field]{Krelowski:1993}.  Our version of this method exploits the relatively narrow range of intrinsic near-infrared colors of stars to separate foreground from background stars. Clouds with sufficiently high column density will redden background stars to such an extent that the intrinsically reddest foreground star will be bluer than the intrinsically bluest background star. We can then simply count the number of stars bluer than this color threshold and compare with a Galactic model of the stellar distribution to estimate a distance. This simple method has the advantage that the distribution of stellar colors in the Galactic model need not be precisely correct, only the number density as a function of distance from the Sun. This method is quite similar to the method applied to derive distances to nearby (D $<$ 1 kpc) molecular clouds in \citet{Lombardi:2010} and to more distant clouds in \citet{Kainulainen:2011} for distant Infrared Dark Clouds (IRDCs).

\begin{figure}
\plotone{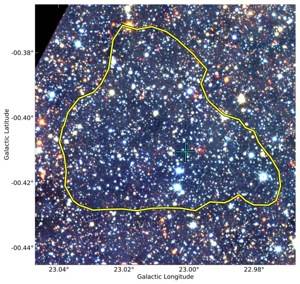}
\caption{UKIDSS three-color image (blue = $J$, green = $H$, red=$K$) of G23.01$-$0.41. The yellow contour shows the boundary of this cloud as determined from the BGPS thermal emission map. The cyan cross marks the position of the maser which has been used to measure a parallax distance toward this cloud. Note that inside the contour the stellar colors are typically blue or very red, as confirmed by the $J-K$ color histogram for this cloud (Figure~\ref{SampleHist}).}
\label{SampleImage}
\end{figure}

\begin{figure}
\plotone{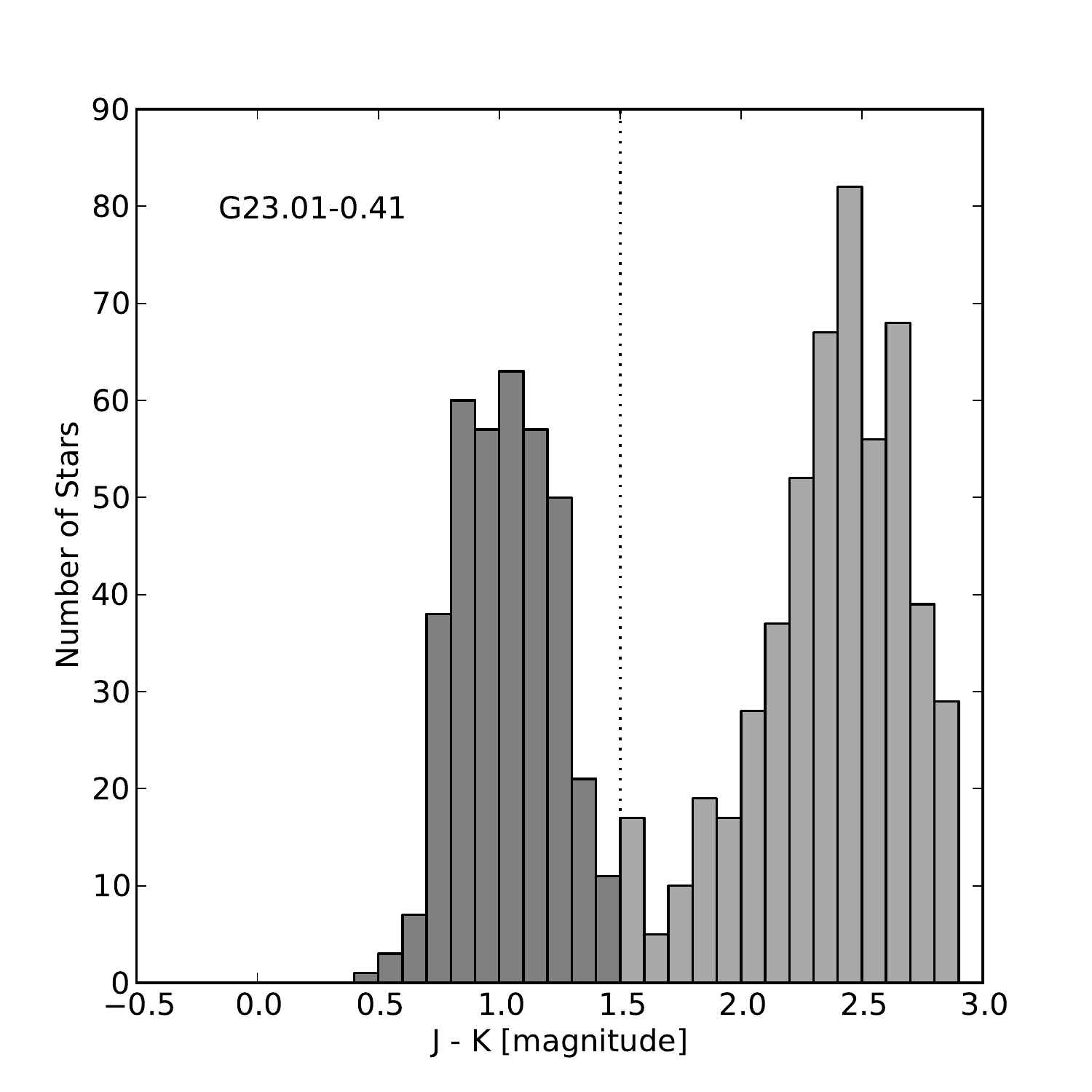}
\caption{$J-K$ color histogram for stars within the contour defined for G23.01$-$0.41 using UKIDSS data. This cloud produces a very clear separation between blue foreground stars ($J-K <$ 1.5 mags; dark gray) and red background stars ($J-K >$ 1.5 mags; light gray). The results are insensitive to the exact value of the color cut as foreground stars are well separated from background stars.}
\label{SampleHist}
\end{figure}

\begin{figure}
\plotone{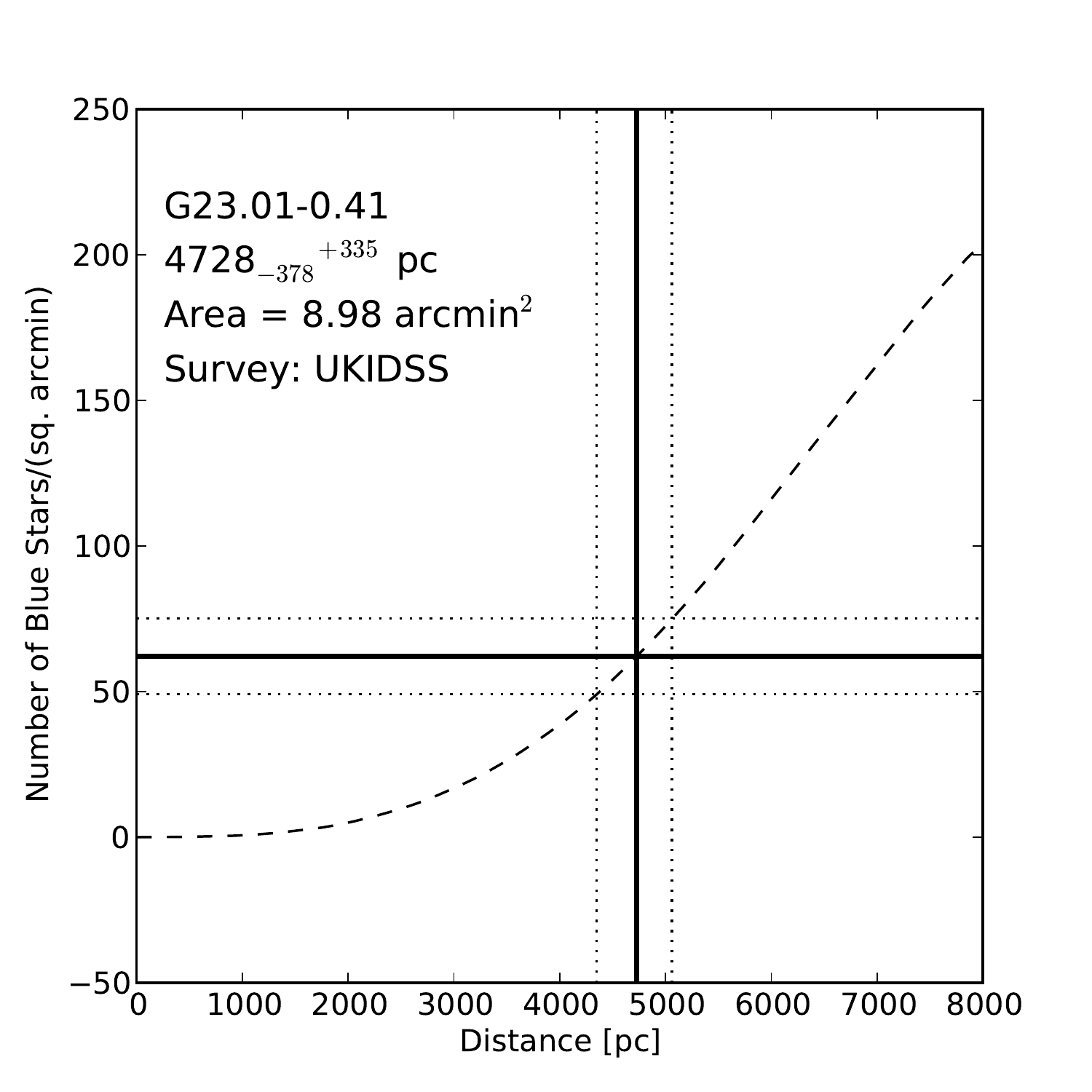}
\caption{Blue Number Count distance to G23.01$-$0.41 using UKIDSS data. We compare the (completeness-corrected) number of blue ($J-K < 1.5$ mags) stars per acrminute within the cloud contour (horizontal line) against the number of stars predicted to be closer than a given distance from the Besan\c con Galactic model of \citet{Robin:2003} (dashed line). Poisson errors on the number of stars produce an estimate of the (non-systematic) uncertainty of this distance determination (vertical lines).}
\label{SampleDetermination}
\end{figure}

The Blue Number Count method involves three steps: (1) define the region of the dark cloud associated with a maser source which is dense enough produce a clear color separation; (2) count the number of blue stars within this region, accounting for confusion and incompleteness; (3) compare the number of blue stars with a Galactic model to determine a distance to the cloud. In the following sections we elaborate on each of these steps and illustrate the method in Figure~\ref{SampleImage}, \ref{SampleHist}, and \ref{SampleDetermination} for an example cloud, G23.01-0.41. We also discuss some of the possible sources of bias and error in this method. 

\subsection{Defining a Dense Region}

The first step in the Blue Number Count method is to identify the column density necessary to make a clear separation between foreground and background stars. The range of intrinsic $J-K$ colors is less than 1.5 magnitudes \citep[e.g.][]{Lombardi:2001}. Therefore we need enough column density to produce at least 1.5 magnitudes of reddening in $J-K$. Using the \citet{Rieke:1985} extinction curve $A_J/A_V = 0.282$ and  $A_K/A_V = 0.112$, thus $A_V = \Delta(J -K)/0.170$ and the extinction required to produce the requisite amount of reddening to clearly separate foreground from background stars is 8.8 magnitudes of A$_V$. 

We use Bolocam Galactic Plane Survey \citep[BGPS;][]{Aguirre:2011} data to identify the boundaries of the cloud with column densities large enough to produce an extinction of at least 8.8 magnitudes of A$_V$. We assume that 
\begin{equation}
N_{H_2} = \frac{S_{\nu}^{beam}}{\Omega_{A} \mu_{H_2} m_H \kappa_{\nu} B_{\nu} (T)},
\end{equation}
where $S_{\nu}^{beam}$ is the flux per beam, $\Omega_{A}$ is the beam solid angle, $\mu_{\mathrm{H_2}}$ is the mass per hydrogen molecule, $\kappa_{\nu}$ is the dust opacity, and $B_{\nu} (T)$ is the Planck function. For Bolocam data, $\lambda = 1120\, \mu$m, and $\theta_{HPBW} =$ 31\arcsec. We used $\mu_{\mathrm{H_2}} = 2.8$, and $\kappa_{\nu} = 0.0114 $ cm$^{2}$ g$^{-1} $ corresponding to the wavelength-interpolated \citet{Ossenkopf:1994} value for dust grains with thin ice mantels in dense regions (n = 10$^{6}$ cm$^{-3}$) and assumed a dust temperature of T = 10 K. For these assumptions
\begin{equation}
N_{H_2}/S_{\nu}^{beam} = 6.77 \times 10^{22} cm^{-2} (\mathrm{Jy \: beam}^{-1})^{-1}
\end{equation}
in the appropriate units (since BGPS data is available in maps with units of Jy beam$^{-1}$). N(H$_2$) is then converted to $A_V$ assuming 
\begin{equation}
N_{H_2} = 9.4 \times 10^{20} cm^{-2} (A_V/mag)
\end{equation}
from \citet{Bohlin:1978}. Therefore, our critical flux threshold of 8.8 magnitudes corresponds to a BGPS flux threshold of 0.12 mJy beam$^{-1}$. The various conversion factors assumed here ($T$, $\kappa_{\nu}$, and the relation between $A_V$ and N($H_2$)) are relatively uncertain, but the exact value we obtain is not critical to our analysis; all we requires is that the adopted flux threshold produce a separation between foreground and background stars. Figure~\ref{SampleImage} displays this 0.12 mJy beam$^{-1}$ contour for our example cloud (G23.01$-$0.41) and shows how this contour matches the area over which stars are either very blue (in front of the cloud) or very red (behind a large column of dust). 

\begin{figure}
\plotone{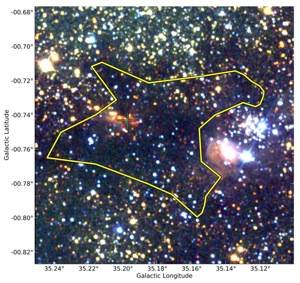}
\caption{Contour for G35.20$-$0.74 made from the GRS $^{13}$CO integrated intensity image overlain on the 2MASS three color image ($J$=blue, $H$=green, $K$=red). The contour was adjusted by hand to exclude the bright blue cluster of stars at $l$=35.12\arcdeg, $b$=-0.75\arcdeg. This cluster may have been born in this cloud, and its exclusion is necessary so that the distance determination is not biased. Note that embedded young stars such as those around the maser (cyan cross) will not bias the Blue Number Count method, since red stars are not used in the analysis.}
\label{AdjustContours}
\end{figure}

Three of our regions are not included in BGPS because their Galactic latitudes were too large to be included in the BGPS. For two of these sources (IRAS 19213+1723, G35.20-0.74) we use the GRS $^{13}$CO data, integrated over $\pm$ 15 km s$^{-1}$ and convert this to a column density using the conversion factors in \citet{Carpenter:2000}. For G35.20-1.74, no GRS or BGPS data is available, and we estimate the A$_V$=8.8 mag region by hand, tracing the obvious dark extinction area. A few other cloud boundaries are adjusted to exclude regions of particularly bright emission or obvious foreground clusters of blue stars (see Figure~\ref{AdjustContours}). 

\subsection{Counting Blue Stars}

\begin{figure*}
\plottwo{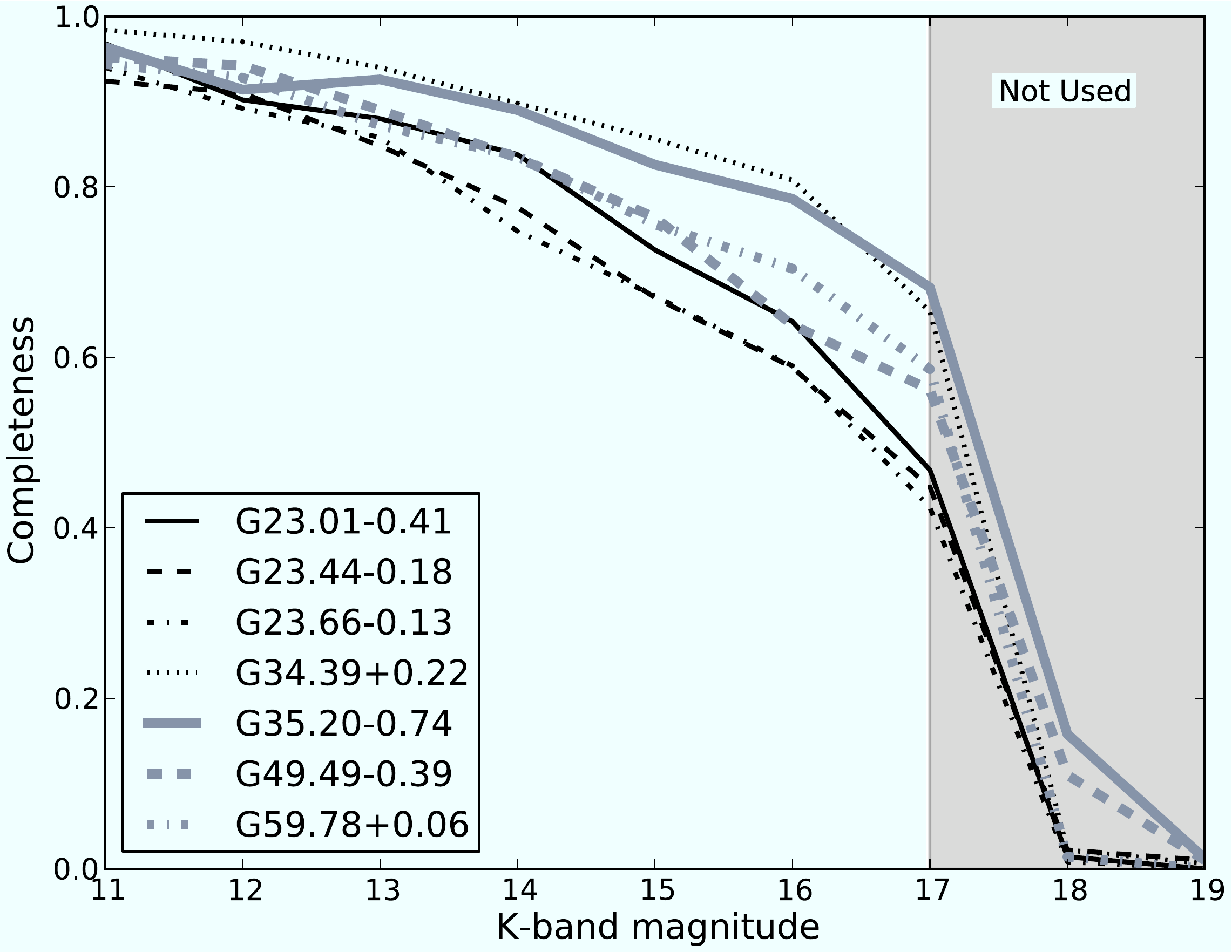}{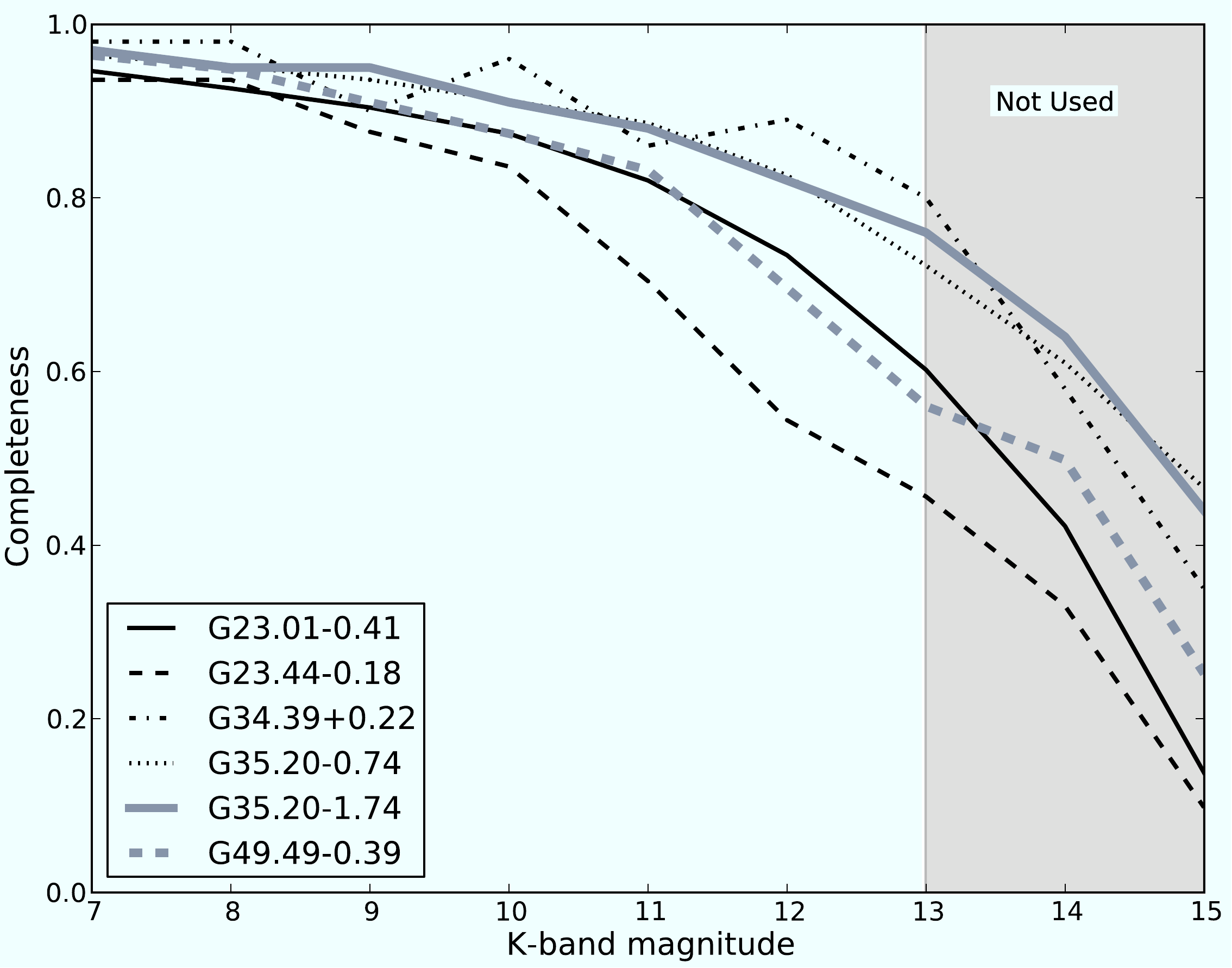}
\caption{$K$-band completeness estimates for our sample of clouds containing masers for [left] UKIDSS and [right] 2MASS. Completeness was estimated by injection and recovery of 500 artificial stars at each integer magnitude. All artificial stars were placed randomly within the cloud contour. Due to the steep drop-off in completeness, we limited our analysis to stars brighter than 17 magnitudes in UKIDSS and 13 magnitudes in 2MASS.}
\label{completeness}
\end{figure*}

The second step in the Blue Number Count method is to count the number of blue stars within the region of interest while correctly accounting for completeness. The masers reside in regions of high extinction and active star formation. Active star formation can produce significant nebulosity and thus a spatially variable background. On the other hand, the high extinction means that confusion is less severe when performing photometry on a dark cloud. Average completeness estimates are therefore unreliable; the completeness of a given survey is different for each cloud.

We performed our own estimates of the completeness by injecting synthetic stars at random positions within the A$_V$=8.8 mag cloud contour and testing for their recovery using automatic photometry with Source Extractor \citep{Bertin:1996}.  For example, to estimate the completeness of a UKIDSS field we set the saturation level at 40,000 ADU and used a 5-pixel (2\arcsec) radius photometric aperture. This size aperture corresponds to the default aperture size {\sc apermag3} in the UKIDSS catalog. We ran 500 trials on each cloud, injecting stars as artificial circular gaussians into the $K$-band UKIDSS image. We estimate the median FWHM from all the stars in the image and use this value for the FWHM for our artificial stars. The noise comes from sitting on top of the noisy background of the image. In each trial we inserted stars with uniform integer magnitudes between 11 and 19 magnitudes, inclusive. We tested whether we retrieved input stars by searching for output stars within 1 pixel of the input position with a magnitude difference of no more than 0.5 magnitudes plus the magnitude uncertainty reported by Source Extractor. In addition we required a Source Extractor {\sc flag} value of less than 4, eliminating stars with saturated pixels or truncated objects but allowing the star to be partially blended with another star. 

We estimated the completeness of the field at integer magnitudes and interpolated between these values in order to weight each measured star appropriately, as shown in Figure~\ref{completeness}. Because the completeness approaches zero beyond 17 magnitudes in $K$, we limit the stars considered to be between $K$ of 11 and 17 magnitudes in order to avoid applying very large correction factor. We estimate completeness from the $K$-band image since the method only uses the blue stars ($J-K$ $<$ 1.5 mags), and the $J$ image is complete to significantly fainter magnitudes. A similar procedure was carried out for 2MASS images; for 2MASS we limited our analysis to 7 mag $<$ $K$ $<$ 13 mag. Our estimate of the completeness in the 2MASS images is brighter than the quoted survey completeness for uncrowded regions of 2MASS \citep[14.3 magnitudes in $K$;][]{Skrutskie:2006} because (1) even with dark clouds blocking out many background stars there are enough foreground stars in the Galactic plane to cause some confusion, (2) there is significant near-infrared nebulosity in many regions, and (3) our parameters for identifying sources and performing photometry may be less sensitive and tolerant of source-blending than the procedures used to generate the 2MASS catalog. 

We can not use the standard point-source catalogs from UKIDSS or 2MASS because our completeness estimate is only appropriate for our particular photometric parameters. Instead, we create our own catalog with the same Source Extractor parameters and quality checks as used in the completeness study and fit these measured magnitudes against UKIDSS and 2MASS magnitudes to derive magnitude zero-points which puts our photometry roughly on the UKIDSS and 2MASS photometric systems. Because our results do not rely on exact colors, small color or magnitude offsets with respect to the standard catalog magnitudes are unimportant. Figure~\ref{SampleHist} shows the $J-K$ stellar color histogram for our example cloud (G23.01$-$0.41). The color separation between foreground and background stars is quite pronounced.

\subsection{Comparing to a Galactic Model}

The final step in the Blue Number Count method is to compare our count of foreground stars with a Galactic stellar distribution model. We use the Besan\c con model \citep{Robin:2003} to generate a synthetic stellar catalog consisting of distances and near-infrared magnitudes for each of our target fields. The Besan\c con model allows the generation of a ``small-field'' catalog of arbitrary size. That is, we can obtain an arbitrarily large sample of stars at the desired position to reduce Poisson noise in the simulated catalog. We assume that gradients in the stellar distribution across our fields are negligible. We include all spectral types and populations available in the model and initially include no diffuse extinction. The model output is a catalog of stars containing both intrinsic properties (i.e. distance) and observed quantities (i.e. near-infrared magnitudes). 

We process the catalog produced from the Besan\c con model to include diffuse Galactic extinction. We use the default model value of 0.7 magnitudes of A$_V$ per kpc. Since typical maser distances are between 1 and 5 kpc, this gives us less than 3.5 magnitudes of visual extinction from the diffuse component which translates into a maximum $E(J-K)$ of 0.57 magnitudes. Our method does not work if there is a foreground cloud which produces significant excess extinction and our method should only be applied to clouds where auxiliary information confirms this exclusion. Small changes to the diffuse extinction with distance will produce a small change in the distance estimate for nearby clouds and a moderate change for distant clouds. If the true diffuse extinction is larger than we assume, then some foreground stars could be reddened so much that we incorrectly identify them as background stars. The exclusion of these foreground stars would make clouds appear closer than they really are; this is of particular concern for distant clouds where the impact of diffuse extinction is more significant. 

An alternative approach to dealing with diffuse extinction is adopted by \citet{Kainulainen:2011}, who use the 3D extinction profiles of \citet{Marshall:2006} to model the extinction as a function of distance along each line of sight. The \citet{Marshall:2006} profiles are derived using the same Galactic model \citep{Robin:2003} and near-infrared data (2MASS), and it is therefore difficult to disentangle the uncertainty in these profiles from the other uncertainties in the model. By focusing only on the very dense portions of clouds we are able to reduce our sensitivity to the exact value of the diffuse extinction at the expense of using a smaller region of the cloud for our analysis. As long as there is a distinct break in the colors of stars we can separate foreground from background stars, and our only dependence on the Galactic model is the number density of stars as a function of distance.

We use the model catalog to define a monotonically increasing relationship between the cloud distance and surface number density of foreground (blue: $J-K <$ 1.5 mags) stars per square arcminute. This relationship saturates at some distance due to a combination of diffuse extinction and (depending on Galactic position) the finite size of the Galactic disk, but remains monotonic for all our clouds well past the estimated distance. We solve for the distance to a cloud by finding the distance at which our synthetic curve of blue foreground stars matches our estimated true number of blue stars, after correcting for completeness. We assume that the major source of random (non-systematic) uncertainty is Poisson ($\sqrt{n}$) noise due to our small number of blue foreground stars and use the number of measured blue stars (before correcting for completeness) in this calculation. Figure~\ref{SampleDetermination} shows this distance determination for our sample cloud (G23.01$-$0.41) using UKIDSS.

\section{Red Giant Extinction Distances}
\label{SecRedGiant}
\begin{figure*}
  \begin{center}
    \begin{tabular}{ccc}
      \resizebox{57mm}{!}{\includegraphics[angle=0]{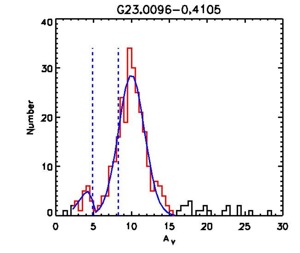}}&
     \resizebox{57mm}{!}{\includegraphics[angle=0]{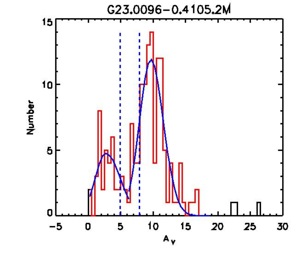}}&
     \resizebox{57mm}{!}{\includegraphics[angle=0]{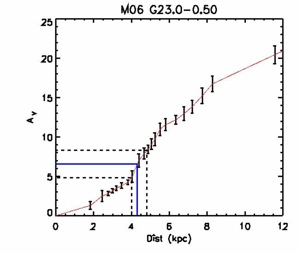}}
    \end{tabular}
    \caption{The Red Giant extinction method for G23.01$-$0.41. [Left; a]: an A$_V$ histogram generated using UKIDSS data extracted from a region containing the maser G23.01-0.41. The near and far sides of the molecular cloud are marked with dashed lines, derived from the 1$\sigma$ deviations of the skewed Gaussians fitted to each histogram (solid curve). [Center; b]: same as (a) except using 2MASS data. [Right; c]: A distance-A$_V$ plot showing the extinction-distance relationship (solid thin line with error bars) created using the \citet{Marshall:2006} data centered at G23.0-0.50. The two sets of dashed lines illustrate how the near and far sides of the molecular cloud, the dashed lines from (a), are converted into a distance and associated error.}
    \label{fig:G23.0096-0.4105_Avhist}
  \end{center}
\end{figure*}

Our second method is the Red Giant extinction distance method. This method makes extinction measurements to individual giant stars along the same line of sight as a molecular cloud. If there are enough foreground and background giant stars available then the presence of the cloud will be apparent in an extinction measurement histogram. An appropriate extinction-distance relationship is then used to convert these measured extinction values into a distance. One of the strengths of this method is that if enough field stars are available then it is possible to detect multiple clouds along the same line of sight.

A full description of this method was presented in \citet{stead10}. However, the most important aspects of the method are also detailed below for completeness. The photometric error cuts for each survey are discussed below. However the UKIDSS data are also subject to further quality control cuts to remove stars with pixels close to saturation, blended objects and other artifacts that may affect the photometry. For further details on the UKIDSS data quality controls please refer to \citet{Lucas:2008}. 

\subsection{Field Star Selection}

The size of the region of sky used to extract field stars depends upon several factors. The area of sky must be large enough to yield an adequate number of extinction measurements to both foreground and background stars. However, the larger the area used, the greater the contamination in the extinction histogram. This contamination occurs predominantly due to the column density structure of the molecular cloud, however overlapping clouds and other irregularities in the interstellar medium will also contribute.

To minimize the contamination in the extinction measurements, where possible we have used GRS data to define the area of sky used for extraction. \citet{rathborne09} identify molecular clouds in the GRS, specifying both their spatial position and their V$_{LSR}$. Integrating the $^{13}$CO luminosity measured in the GRS along a specific range of V$_{LSR}$ will map out the molecular cloud column density. The GRS  $^{13}$CO maps give the ability to select an area of sky containing a similar cloud column density. Isolating these regions will therefore limit the uncertainty caused by cloud patchiness. As previously stated, the exact size of the region should be as large as possible while minimizing the noise in the extinction histogram. The size of each region is decided on a cloud-by-cloud basis after assessment of each extinction histogram.

\subsection{Isolating Red Giants}

Late type giant stars are old enough to have undergone several Galactic rotations. They have therefore moved from their original birth places in the spiral arms, with reference to the molecular clouds, and they can be considered ubiquitous throughout the entire Galaxy. Late type giants are very numerous and are also infrared bright. These qualities, plus the fact that they have known colors, are why many authors have used them to create extinction maps of the Galaxy \citep{Marshall:2006,lombardi09}.

Late type giants occupy well known regions on both color-color and color-magnitude diagrams (CCDs/CMDs). For this reason \citet{stead10} were able to use a series of color cuts to isolate them from the general field population. They tested their method, with realistic synthetic data, to confirm that over 90$\%$ of the final sample of stars will be G0III stars or later. Their method has been replicated for the work in this paper.

\subsection{Extinction Measurements and A$_V$ Histograms}

In the same manner as \citet{stead10}, the color-selected sample of late type giants have been de-reddened, in color-color space, using a K0III reddening track to their point of intersection with the giant locus. This reddening track has been created using a K0III stellar model from \citet{castelli04} and the average extinction power law of \citet[][$\alpha$=2.14]{stead09} using the method described therein. 

The length of the reddening track traversed during the de-reddening process relates directly to the amount of extinction each individual star suffers. Photometric errors, as well as directly adding to the error in the extinction measurement, will also make the point of intersection with the giant locus more uncertain. Photometric errors are the greatest source of noise in an extinction histogram. The faintest late-type giant stars in each region, i.e. those with the largest photometric errors, will be those stars that are behind each molecular cloud. As such stars are essential to the detection of each molecular cloud there is therefore a tradeoff between completeness and noise in the extinction histogram. For the purpose of this paper we consider only stars with photometric errors of less than 0.05 mags in the UKIDSS catalog and less than 0.10 mags in the 2MASS catalog. The less stringent error limit on the 2MASS catalog is used to provide a sufficient number of stars.

A value of A$_J$ is determined directly for each individual star using this de-reddening process. This value is then scaled to a value of A$_V$ using the ratios A$_J$/A$_V$=0.2833 and A$_J$/A$_V$=0.2899 for UKIDSS and 2MASS respectively. These have been calculated using the \citet{cardelli89} extinction curve with R$_V$=3.1. As noted in the literature, the value R$_V$=3.1 may not be appropriate for molecular clouds, and our extinction is likely produced by a combination of diffuse extinction and extinction from molecular clouds. Adopting an extinction law is necessary, since we must compare our derived values of  A$_J$ with the extinction distance relations of \citet{Marshall:2006} which are in A$_{K_S}$. However, the uncertainty introduced by our choice of R$_V$ is small. The conversion between A$_J$ and A$_{K_S}$ differs by only 1.3\% if we choose R$_V$ = 5.5 instead of R$_V$ = 3.1. Note that since our choice of R$_V$ simply determines the multiplicative conversion factors between the different wavelengths, the fact that we are converting A$_J$ and A$_{K_S}$ to A$_V$ before making the comparison does not introduce any additional uncertainty.  

We present an example of this method using the same cloud (G23.01$-$0.41) as was used to illustrated the Blue Number Count method (Figure~\ref{fig:G23.0096-0.4105_Avhist}). The cloud that contains the maser was mapped in $^{13}$CO in the GRS. Stars have been selected from within the $>$8 K km s$^{-1}$ GRS $^{13}$CO integrated intensity contour. As mentioned previously, this value was selected after inspection of the A$_V$ histogram to optimize the completeness/noise tradeoff. Using the conversion factors in \citet{Carpenter:2000} this corresponds to an A$_{V}$ of 4.1 magnitudes, and is thus a lower column density threshold than used in the Blue Number Count method.

Figure~\ref{fig:G23.0096-0.4105_Avhist} (a) displays an A$_V$ histogram created using UKIDSS data. The histogram begins with a small rise in star counts, peaking at A$_V$$\sim$3, and then quickly dips again to a minimum at A$_V$$\sim$5. Following this dip the star counts quickly rise again to form a second peak at  A$_V$$\sim$10. This double peaked histogram is consistent with the idea that there is a single molecular cloud along the line of sight. \citet{stead10} define the gap between the two peaks, measured by fitting skewed Gaussians to each peak to determine the respective near and far 1$\sigma$ widths, as the A$_V$ of the cloud. They used synthetic data to determine the error in this measurement to be A$_V$$\pm$0.5 mag. This value has been included in all of the distance error assessment that follows.

Figure~\ref{fig:G23.0096-0.4105_Avhist} (b) shows the same A$_V$ histogram for 2MASS data. The two obvious differences between each histogram are the higher star counts in the deeper UKIDSS data and the more jagged 2MASS histogram, owing to the smaller number of stars and the less precise 2MASS photometry. 

\subsection{Distance-A$_V$ plots}

\citet{Marshall:2006} used the $\Bes$ Galactic model and 2MASS data to generate a three dimensional map of the Galactic interstellar extinction distribution. They gridded the Galaxy into 15$^{\prime}$x15$^{\prime}$ tiles and generated an extinction-distance relationship (EDR) for each tile. The line of sight extinction measurements in this paper are converted to distances using the \citet{Marshall:2006} EDRs. The spatially closest EDR to each cloud has been used.

The \citet{Marshall:2006} data are presented in terms of distance vs. A$_{K_S}$, the extinction in the 2MASS $K_S$ filter. Figure~\ref{fig:G23.0096-0.4105_Avhist} (c) contains the EDR centered at G23.0-0.50. The UKIDSS measurements produce an estimate for the distance to the cloud of D = 4.3$^{+0.5}_{-0.3}$ kpc. This process is also repeated using 2MASS data yielding a distance, D = 4.3$^{+0.4}_{-0.3}$ kpc. Typically the errors produced using 2MASS data are larger than those derived using UKIDSS data. For this cloud however the 2MASS data have constrained the result somewhat more reliably than with UKIDSS data. The reasons for this are discussed below.

\subsection{UKIDSS versus 2MASS}
\label{subsection:ukidssversus2mass}
As previously discussed, an adequate number of both foreground and background stars are required to make the Red Giant method work. For this reason if a cloud is too close, and there are too few foreground stars available, it will not be possible to constrain a distance to the cloud. Likewise, if the photometry is not deep enough, too few background stars will be detected. However whereas the ability to derive distances using 2MASS data depends only upon photometric depth, the UKIDSS data have a completeness issue at the brighter end caused by the saturation limit. Stars in UKIDSS typically saturate at $K$$\sim$11 mag when selecting only the most reliable data. Therefore, for nearby clouds most foreground giant stars may be saturated and thus unavailable in UKIDSS. UKIDSS and 2MASS complement each other well in this respect, with 2MASS able to constrain distances to nearby clouds and UKIDSS able to probe to much further distances. In the above example, towards this particular region of the Galaxy, $\sim$4 kpc is the point where 2MASS outperforms UKIDSS. This is apparent in Fig. \ref{fig:G23.0096-0.4105_Avhist} (a) and (b) when a comparison is made between the number of foreground stars in each histogram. A merged catalog containing 2MASS and UKIDSS data for the bright and faint stars, respectively, would be ideal, but this is beyond the scope of this work.

\begin{deluxetable*}{lcccccccc}[ht]
\tablecaption{Distances to All Clouds}
	\tabletypesize{\scriptsize}
	\tablehead{
	\colhead{Name} & \colhead{Maser Parallax} &
	\multicolumn{2}{c}{Blue Number Count Extinction}  &
	\multicolumn{2}{c}{Red Giant Extinction}  &
	\multicolumn{2}{c}{Kinematic}  \\
	&  & \colhead{2MASS} & \colhead{UKIDSS}
	& \colhead{2MASS} & \colhead{UKIDSS}
	& \colhead{Clemens (1985)} & \colhead{Reid et al. (2009)}
	}
	\startdata
	G05.89$-$0.39		&	$1.28_{-0.08}^{+0.09}$		&	\nodata				&	$4.5_{-0.8}^{+0.6}$	&	$3.9^{+1.4}_{-1.1}$	&	$3.7^{+0.3}_{-0.4}$	&	$2.0_{-0.7}^{+0.7}$	&	$1.9_{-0.7}^{+0.6}$	\\[2mm]
					&			\nodata			&	\nodata				&		\nodata		&	$6.9^{+0.4}_{-0.4}$	&	$6.2^{+0.3}_{-0.3}$	&		\nodata		&		\nodata		\\[2mm]
	G09.62$+$0.20	&	$5.20_{-0.6}^{+0.6}$			&	\nodata				&	$2.6_{-1.1}^{+0.6}$	&	$4.4^{+0.4}_{-0.3}$	&	$6.2^{+1.2}_{-1.1}$	&	$0.9_{-0.5}^{+0.5}$	&	$0.9_{-0.6}^{+0.5}$	\\[2mm]
					&			\nodata			&	\nodata				&		\nodata		&	$7.4^{+0.7}_{-0.7}$	&	\nodata		&		\nodata		&		\nodata		\\[2mm]
	G23.01$-$0.41		&	$4.59_{-0.33}^{+0.38}$		&	$4.0_{-1.4}^{+0.9}$		&	$4.7_{-0.4}^{+0.3}$	&	$4.3^{+0.4}_{-0.3}$	&	$4.3^{+0.5}_{-0.3}$	&	$5.0_{-0.1}^{+0.1}$	&	$4.6_{-0.1}^{+0.1}$	\\[2mm]
	G23.44$-$0.18		&	$5.88_{-0.93}^{+1.37}$		&	$5.1_{-0.7}^{+0.6}$		&	$4.8_{-0.2}^{+0.2}$	&	$5.6^{+0.3}_{-0.8}$	&	$4.8^{+1.2}_{-1.7}$	&	$6.3_{-0.2}^{+0.2}$ 	&	$5.4_{-0.1}^{+0.1}$	\\[2mm]
	G23.66$-$0.13		&	$3.19_{-0.35}^{+0.46}$		&	\nodata				&	$4.9_{-0.2}^{+1.3}$	&	\nodata			&	\nodata			&	$5.2_{-0.1}^{+0.1}$	&	$4.7_{-0.1}^{+0.1}$	\\[2mm]
	G34.39$+$0.22	&	$1.56_{-0.11}^{+0.12}$		&	$2.9_{-1.2}^{+0.9}$		&	$3.6_{-0.4}^{+0.3}$	&	$2.6^{+0.2}_{-0.5}$	&	$2.6^{+0.2}_{-0.4}$	&	$3.7_{-0.2}^{+0.2}$	&	$3.5_{-0.2}^{+0.2}$	\\[2mm]
					&			\nodata			&		\nodata			&		\nodata		&	$3.3^{+0.6}_{-0.2}$	&	$3.6^{+0.8}_{-0.5}$	&		\nodata		&		\nodata		\\[2mm]
	G35.20$-$0.74		&	$2.19_{-0.20}^{+0.24}$		&	$2.4_{-0.7}^{+0.8}$		&	$2.7_{-0.3}^{+0.3}$	&	$2.4^{+0.3}_{-0.2}$	&	$2.6^{+0.7}_{-0.2}$	&	$2.3_{-0.2}^{+0.2}$	&	$2.4_{-0.2}^{+0.2}$	\\[2mm]
	G35.20$-$1.74		&	$3.27_{-0.42}^{+0.56}$		&	$2.5_{-1.0}^{+0.7}$		&	\nodata			&	\nodata			&	$2.9^{+0.2}_{-0.4}$	&	$2.8_{-0.2}^{+0.2}$	&	$2.8_{-0.2}^{+0.2}$	\\[2mm]
	W51 Main/S		&	$5.41_{-0.20}^{+0.21}$		&	$4.0_{-1.2}^{+1.2}$		&	$6.1_{-0.6}^{+0.6}$	&	\nodata			&	$5.9^{+1.1}_{-1.7}$	&	$5.5_{-1.1}^{+1.1}$	&	$5.5_{-0.8}^{+0.8}$	\\[2mm]
	IRAS 19213$+$1723 &	$3.98_{-0.50}^{+0.67}$		&	\nodata				&	\nodata			&	$2.9^{+0.5}_{-0.8}$	&	$4.6^{+2.2}_{-0.8}$	&	$3.5_{-0.3}^{+0.4}$	&	$3.8_{-0.4}^{+0.5}$	\\[2mm]
	G59.78$+$0.06	&	$2.16_{-0.09}^{+0.10}$		&	\nodata				&	$3.8_{-1.2}^{+0.9}$	&	$2.5^{+0.2}_{-0.2}$	&	\nodata			&	$3.3_{-0.5}^{+1.0}$	&	$4.2_{-1.0}^{+1.0}$	
	\enddata
	\tablecomments{All distances are in kpc}
	\label{Table:Distances}
\end{deluxetable*}

\section{Comparison of Methods in 1st Quadrant Sources}
\label{SecComparison}

Table~\ref{Table:Distances} lists the distance determinations for all methods. We do not report a distance measurement for some combinations of sources and methods, either because the method fails to identify a cloud or because certain data are unavailable. For instance, we do not report a Blue Number Count method distance if there are fewer than five blue foreground stars. For clouds with small angular extent there are often not enough blue 2MASS stars to make a measurement. For G35.20$-$1.74 UKIDSS data images were not available, although catalog information was available. Therefore, the Red Giant method (which uses the catalog) was able to determine a distance for this source using UKIDSS data, but the Blue Number Count method (which requires the image data to estimate completeness) was not able to make a determination using UKIDSS. As discussed in \S\ref{subsection:ukidssversus2mass}, the Red Giant method requires foreground and background red giants, so for some clouds too many of the foreground giants are saturated in UKIDSS; conversely, in some clouds there are insufficient background red giants detected in the less sensitive 2MASS catalog.

\begin{figure*}
\plotone{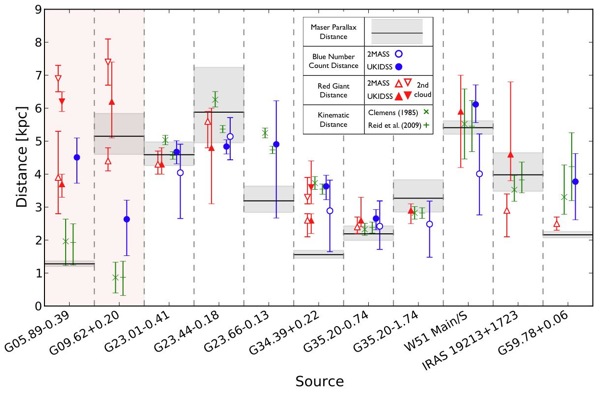}
\caption{Comparison of different distance estimates to the maser clouds, sorted by Galactic longitude.  Black lines and gray shaded regions show the maser parallax distance 1$\sigma$ uncertainties taken from the literature. Blue points show Blue Number Count distances using 2MASS (open symbols) and UKIDSS (filled symbols). Red points show Red Giant distances using 2MASS (open symbols) and UKIDSS (filled symbols); this method sometimes identifies multiple clouds along the line of sight, in which case both distances are shown. Green points show kinematic distances based on two different rotation curves. All error bars are 1$\sigma$ and are estimated for each method as explained in the text. Red shaded clouds show the two sources near the Galactic center where kinematic distances are unreliable and there is evidence for multiple clouds along the line of sight. Figure~\ref{AllDistancesSortedByDistance} shows this figure sorted by maser parallax distance.}
\label{AllDistances}
\end{figure*}

\begin{figure*}
\plotone{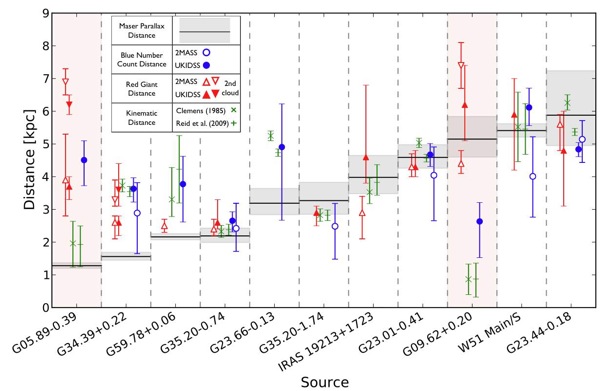}
\caption{Comparison of different distance estimates to the maser clouds. This is the same as Figure~\ref{AllDistances} except sorted by increasing maser parallax distance. Red shaded clouds show the two sources near the Galactic center where kinematic distances are unreliable and there is evidence for multiple clouds along the line of sight. The non-parallax methods (both kinematic and extinction) appear to be systematically biased toward deriving larger distances for near clouds ($D < $ 2.5 kpc). }
\label{AllDistancesSortedByDistance}
\end{figure*}

Figure~\ref{AllDistances} and Figure~\ref{AllDistancesSortedByDistance} compare all our methods for our sample of 1st quadrant masers sorted by Galactic longitude and maser parallax distance, respectively. The two sources near the Galactic center are shaded red in both these figures. For these sources kinematic distances are unreliable and there is strong evidence for at least two clouds along the line of sight (see \S\ref{subsec:galacticcenter} for more details). As expected, the different distance methods often disagree for these two clouds. Table~\ref{comparison} shows the absolute number and fraction of sources for which each method is within 2$\sigma$ agreement of the maser parallax distance (using the $\sigma$ for each method added in quadrature). As this table shows, the methods generally agree within their errors.

There are a few notable exceptions where the methods fail. These clouds are G23.66$-$0.13, G34.39$+$0.22, and the two clouds near the Galactic center (G05.89$-$0.39 and G09.62$+$0.20). We discuss these clouds individually in \S~\ref{subsec:comments}. For the rest of the sources the agreement is generally good. In particular there are clouds (such as G35.20$-$0.74 and G23.01$-$0.41) where all the distances agree extremely well and have small uncertainty ($\sigma$$<$0.5 kpc).

Figure~\ref{AllDistancesSortedByDistance} suggests that there is a systematic tendency to overestimate the extinction distances for nearby sources. In particular, G05.89$-$0.39, G34.39$+$0.22, and (to a lesser extent) G59.78$+$0.06 all have maser parallax distances $<$ 2.5 kpc and kinematic/extinction distances which are systematically larger. Nearby clouds can present particular challenges for both extinction methods. If all foreground giants are saturated, then the Red Giant method will fail to see a clean separation. Very nearby clouds also have a low density of foreground blue stars for the Blue Number Count method, although this is often offset by the fact that nearby clouds are larger in angular extent.

\begin{deluxetable}{lll}
\tablecaption{2$\sigma$ Agreement with Maser Distances}
	\tabletypesize{\footnotesize}
	\tablehead{
	\colhead{Method} & \colhead{Full Sample} & \colhead{Excluding $l < 15$}}
	\startdata
	Blue Number Count (2MASS) & 6/6 (100\%) & 6/6 (100\%) \\
	Blue Number Count (UKIDSS) & 6/9 (66.7\%) & 6/7 (85.7\%) \\
	Red Giant (2MASS) & 7/8 (87.5\%) & 6/6 (100\%) \\
	Red Giant (UKIDSS) & 7/9 (77.8\%) & 6/7 (85.7\%) \\
	Kinematic (Clemens) & 7/11 (63.6\%) & 6/9 (66.7\%) \\
	Kinematic (Reid) & 8/11 (72.3\%) & 7/9 (77.8\%) \\
	\enddata
	\label{comparison}
\end{deluxetable}

Both extinction distance methods rely on average stellar distributions, so small-scale inhomogeneities in the stellar population can produce erroneous distance estimates. In the Blue Number Count method the reliance on average stellar properties is explicit since we use the average stellar density along a given line of sight; in the Red Giant method this dependence arises through the use of the \citet{Marshall:2006} extinction-distance relationships, which rely on average stellar colors. Small-scale inhomogeneities will produce an incorrect distance in the Blue Number Count method; for instance, a (weak, hard-to-detect) stellar cluster projected onto the line of a dark cloud would cause us to estimate an incorrectly large distance. For more distant clouds, we expect that small-scale inhomogeneities in the foreground population will tend to average out and be less significant. The extinction-distance relations of \citet{Marshall:2006} used in the Red Giant method rely on fitting the full color distribution of stars toward a given position. Since this method considers all stars toward a given line of sight, we would not expect small-scale inhomogeneities to produce a bias for any particular distance.

One possible explanation why both the extinction and the kinematic distances fail for certain clouds would be the presence of a spiral arm. Spiral arms may be associated with shocks and therefore departures from the circular Galactic rotation curve. In addition, spiral arms are not included in the $\Bes$ Galactic model, and so neither the stellar populations in these arms, nor the extinction within them, are properly modeled. The old stellar population (including the red giants) will typically be well mixed, and it is only the young stellar population that contributes to the arm/interarm contrast. Spiral arm tangencies in the first quadrant are around $l$ = 25\arcdeg\ and $l$ = 35\arcdeg \citep{Benjamin:2009}. Unsurprisingly, we have three clouds along each of these lines of sight, as regions of high-mass star formation tend to occur in spiral arms. Most of these clouds have extinction distances that agree with the maser parallax distances, suggesting that looking along spiral arms does not tend to produce erroneous distance estimates.

Finally, it is possible that the maser parallax distances may be incorrect. Although the maser parallax distances require significantly fewer astrophysical assumptions, there are possible situations in which these parallax determinations could fail. The fundamental assumption in these observations is that the masers in these objects persist and move in a regular fashion over the several years when observations are made. This is typically mitigated by the use of multiple maser spots. However,  uncompensated atmospheric delays can produce systematic errors in the parallax measurement. 

There are a few masers for which multiple groups have measured the parallax independently, and in general the agreement is within the estimated uncertainties. However in one case, G48.61+0.02, the two distance measurements are 10$\sigma$ discrepant \citep[][Reid private communication; ]{Nagayama:2011}. This demonstrates that it is possible that some maser parallax distances are incorrect.

\subsection{Comments on Individual Sources}
\label{subsec:comments}

\subsubsection{Galactic Center Sources (G05.89$-$0.39 and G09.62$+$0.20)}
\label{subsec:galacticcenter}

\begin{figure*}
\plottwo{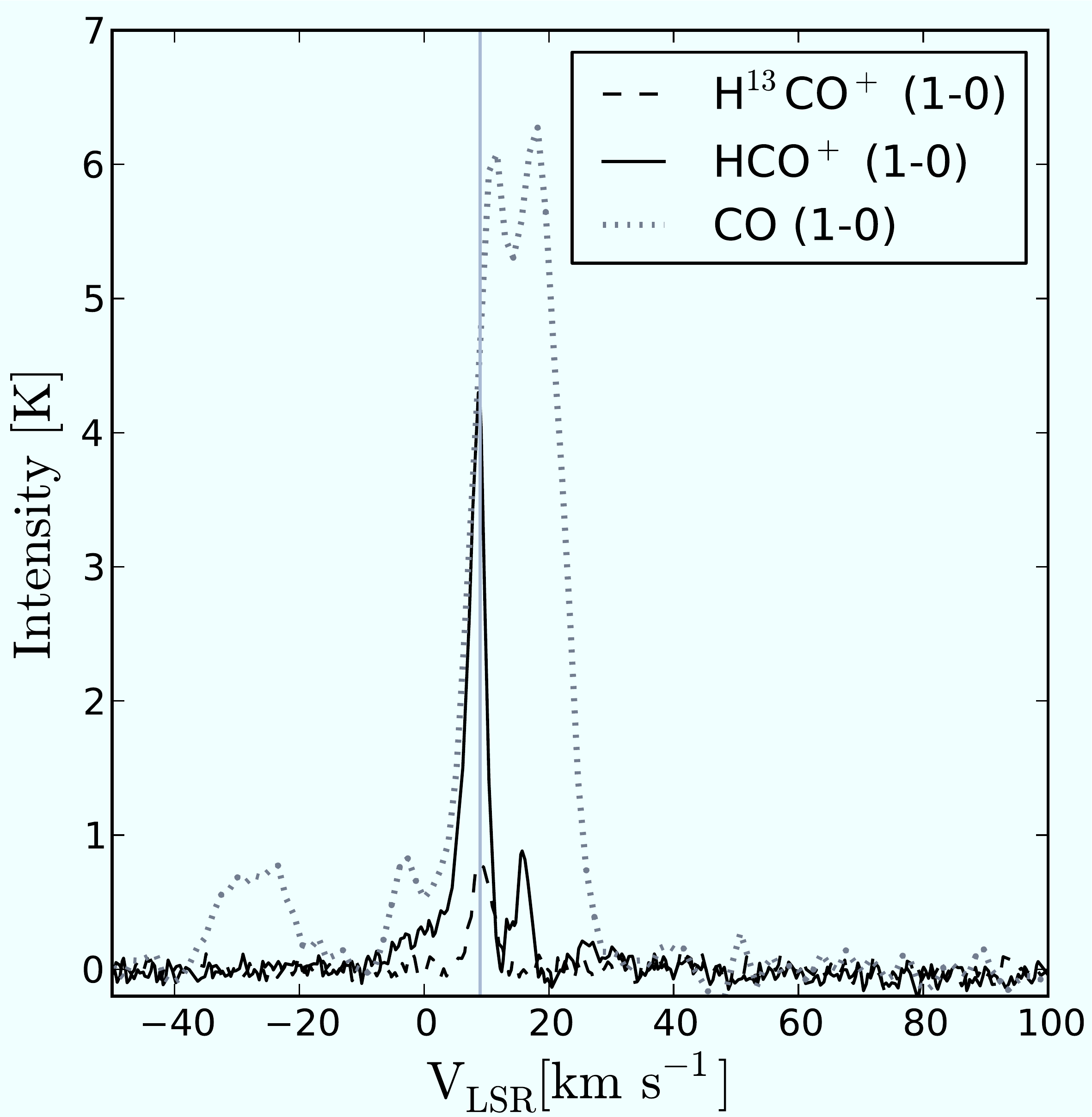}{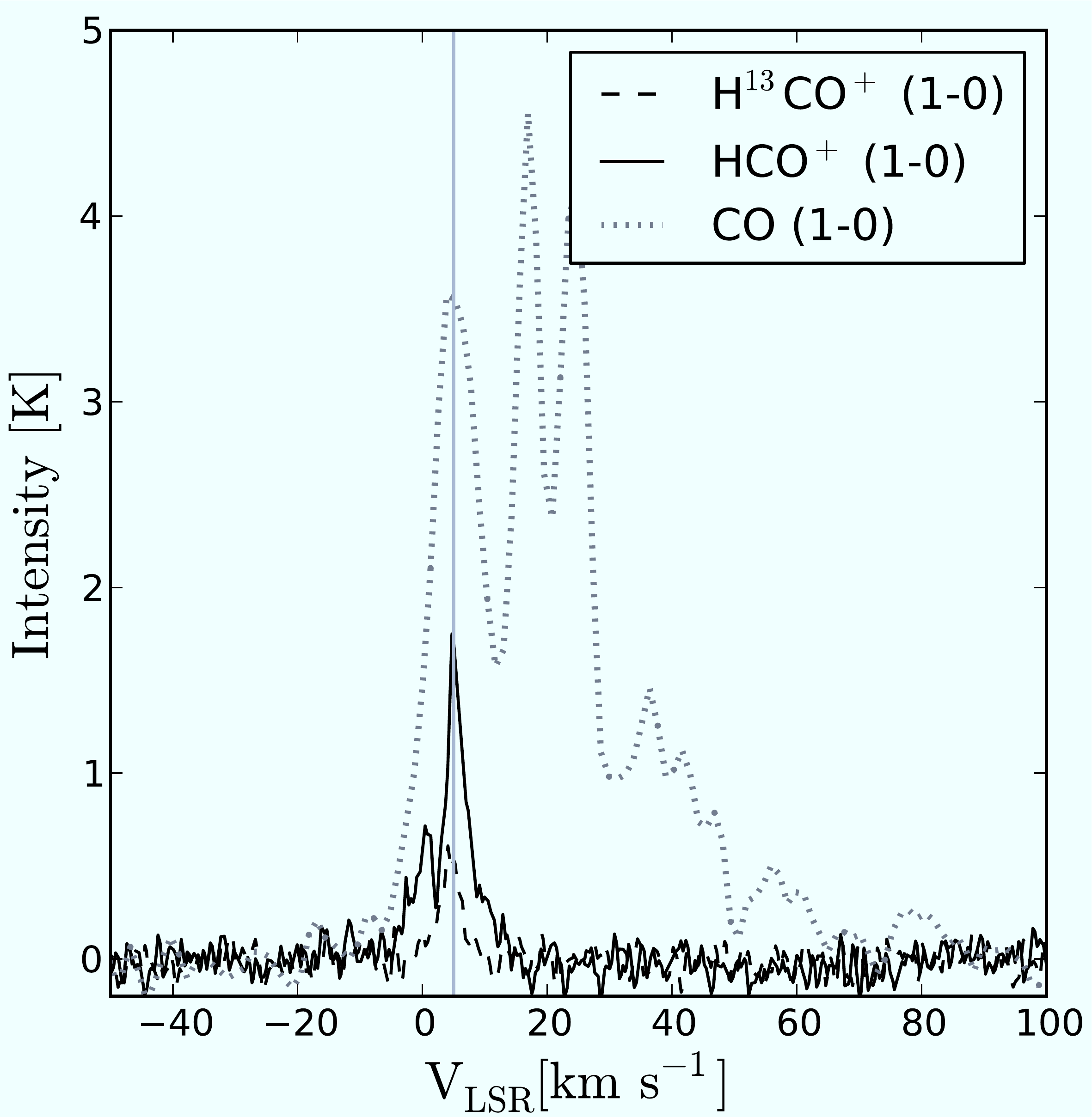}
\caption{Multiple velocity components in dense gas tracers toward the two clouds closest to the Galactic center (right: G05.88$-$00.39, left: G09.62$+$00.20) indicate the presence of multiple clouds along the line of sight, confounding extinction distance determinations here. CO (1-0) spectra are from \citep{Dame:2001} and show an abundance of velocity features toward these positions. The HCO$^+$ (1-0) and H$^{13}$CO$^{+}$ (1-0) spectra come from the MALT90 Survey \citep{Foster:2011}. The high critical density ($>$10$^{5}$ cm$^{-3}$) of these molecules mean they are picking out just dense clouds. Vertical line shows the velocity used for kinematic distances. Both spectra show multiple velocity components in HCO$^+$ (1-0). The optically thin H$^{13}$CO$^{+}$ (1-0) spectrum shows that the multiple velocity features seen in the HCO$^+$ (1-0) spectra are not due to self-absorption.}
\label{MultipleComponents}
\end{figure*}

It is difficult to determine distances for sources that are close to the Galactic center. Kinematic distances are unreliable close to the Galactic center since most orbital motion is perpendicular to the line of sight. Both extinction methods have problems if there are multiple clouds along the line of sight; the projected surface density of molecular clouds, and thus the chance of multiple clouds along the line of sight, increases toward the Galactic center. With sufficient data, the Red Giant method can distinguish multiple jumps in extinction and therefore identify multiple clouds. The Blue Number Count distance method is sensitive only to the first dense cloud along the line of sight. 

For the two sources close to the Galactic center (G05.89$-$0.39 and G09.62$+$0.20) one or more of the methods presented in this paper fails to obtain a distance consistent with the maser parallax distance. The Red Giant method identifies multiple clouds along the line of sight toward both these positions. Spectra from the Millimeter Astronomy Legacy Team 90 GHz (MALT90) survey \citep{Foster:2011} confirm that there are at least two clouds by detecting two velocity components in dense molecular gas tracers (Figure~\ref{MultipleComponents}). 

The combination of multiple clouds and nearly perpendicular orbital motion explains why the methods presented in this paper do not match the maser parallax distances for these two clouds close to the Galactic center. We shade these sources in red in Figure~\ref{AllDistances} and Figure~\ref{AllDistancesSortedByDistance} and analyze them separately in Table~\ref{comparison}. We caution that extinction distances should always be combined with spectra of a dense gas tracer to detect multiple clouds along the line of sight and thereby identify distance estimates which are potentially unreliable.

\subsubsection{G23.66$-$0.13}

This source shows a significant discrepancy between the kinematic distances and the maser parallax distance. Although it is less well characterized as a region of high-mass star formation, there are several lines of evidence in support of this identification. The source was targeted for a methanol maser search based on its Infrared Astronomical Satellite (IRAS) colors; these observations revealed a ring of 6.7 GHz methanol masers \citep{Bartkiewicz:2005}. The discovery of additional objects with similar maser morphologies led \citet{Bartkiewicz:2009} to propose a new class of ``ring-like'' methanol masers. The source is included in the Red MSX Source (RMS) Survey \cite{Urquhart:2011} where it shows an spectral energy distribution consistent with an HII region and has detections of NH$_{3}$ (1,1), (2,2) and (3,3) with V$_{LSR}$ = 80.45 km s$^{-1}$ toward the position.

\begin{figure*}
\plottwo{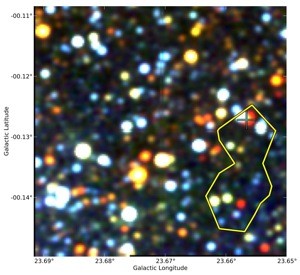}{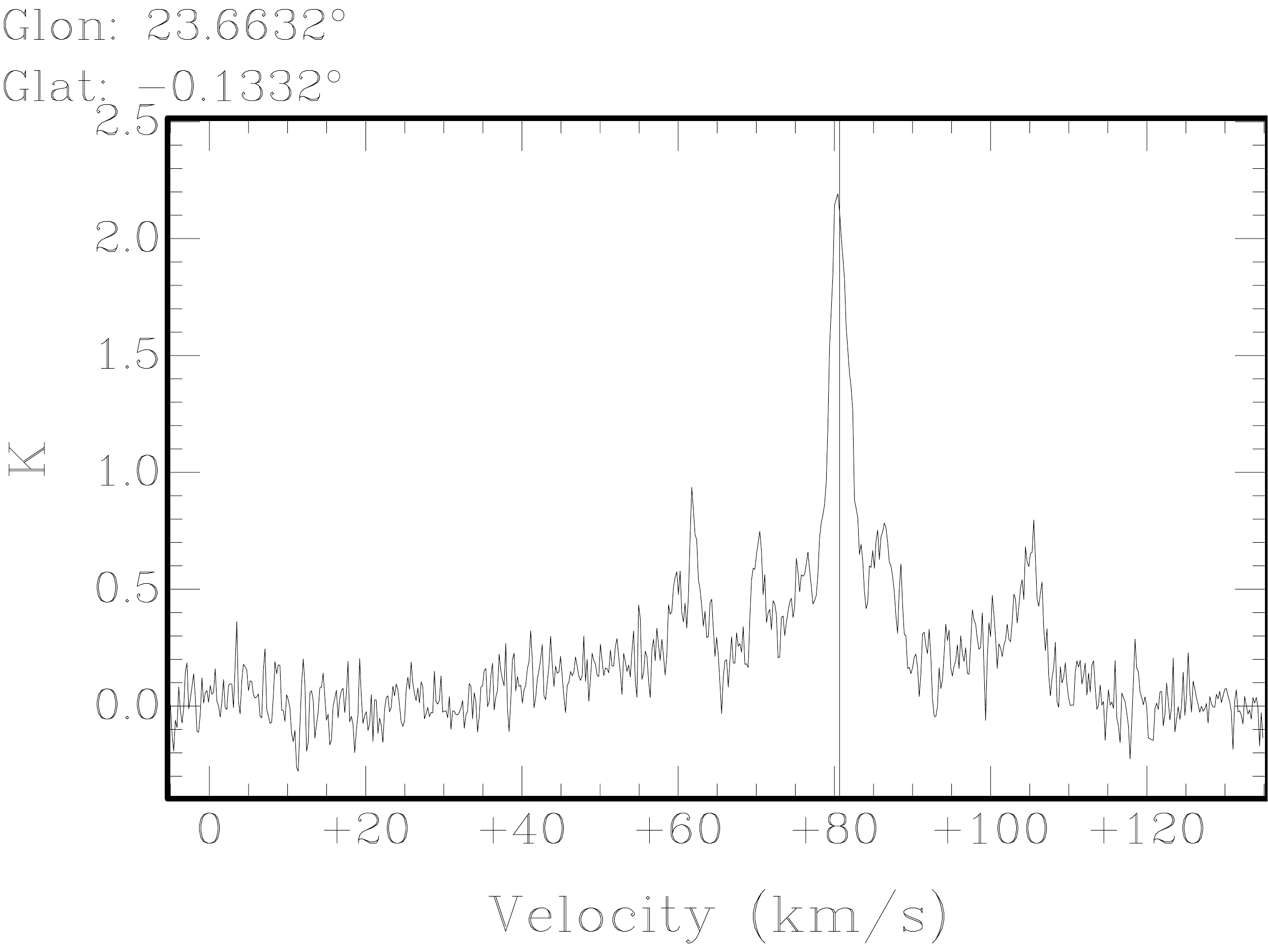}
\caption{[Left] 2MASS three-color image ($J$ = blue, $H$ = green, $K$ = red) of G23.66$-$0.13. BGPS contours at 0.12 mJy beam$^{-1}$ are in yellow, and the position of the maser source is shown by the cyan cross. [Right] GRS  $^{13}$CO (1-0) spectrum at the position of the maser source G23.66-0.13. The strongest emission is at +80 km s$^{-1}$, consistent with the RMS NH$_3$ velocity for this source \citep{Urquhart:2011} and the central velocity of the methanol masers in this source (+82 km s$^{-1}$), confirming the association of the maser source with the cloud identified in BGPS and GRS. }
\label{RingMaser}
\end{figure*}

We detect a small dark cloud associated with this maser position. Figure~\ref{RingMaser} shows the 0.12 mJy beam$^{-1}$ contours of the BGPS data at this cloud overlaid on the 2MASS image. The $^{13}$CO (1-0) spectrum from the GRS of this position shows a single strong line at 80 km s$^{-1}$ consistent with both the NH$_{3}$ velocity and the central maser velocity and spatially consistent with the dark cloud. Because the dark cloud toward this source is so small, extinction distances are difficult. Only the UKIDSS data contains enough blue stars for a (highly uncertain) estimate of the distance from the Blue Number Count method. The Red Giant Method does identify a cloud at 3.2$^{+0.6}_{-0.4}$ kpc by choosing a very small extraction window around the maser position. However, this extinction feature is only 1 magnitude in A$_{V}$, far below our normal column density threshold for identifying a cloud; for this reason we do not consider this distance a reliable estimate.

In conclusion, although G23.66$-$0.13 is not as well-studied a region of high-mass star formation as the other objects considered in this study, it does appear to be associated with a small dense cloud with a velocity consistent with its 6.7 GHz methanol masers, lending credence to the identification of this object as a young high-mass star. The other possible explanation is that the maser parallax distance is measuring the distance to an asymptotic giant branch (AGB) star which is spatially coincident with the dark cloud. AGB stars can power masers, though there is no known case of an AGB star powering a 6.7 GHz methanol maser. We conclude that the AGB explanation is unlikely and the significant difference between the kinematic distance and maser parallax distance remains unexplained.

\subsubsection{G34.39$+$0.22}

For the source G34.39$+$0.22, the extinction distances presented herein have small uncertainties and are consistent with the kinematic distances for this cloud, but are inconsistent with the maser parallax distance (the 2MASS-based Blue  Number Count method is consistent with the maser parallax distance due solely to its large uncertainty). The Red Giant extinction distance method identifies two clouds towards this line of sight, but the first cloud is a rather tenuous identification and spectra of dense gas tracers toward this source \citep[e.g CS (2-1), HCO$^{+}$ (1-0);][]{Sanhueza:2010} do not show a second velocity component. Using either of the rotation curves in this work, the cloud would need to have V$_{LSR}$ $\sim$20 km s$^{-1}$, a discrepancy of 37 km s$^{-1}$, in order for the kinematic distance to match the maser parallax distance. 

There is weak CO emission around 27  km s$^{-1}$ seen in both the GRS $^{13}$CO (1-0) spectra and Atacama Pathfinder Experiment (APEX) $^{13}$CO (3-2) spectra (Sanhueza, private communication), but it seems highly unlikely that the multiple maser spots measured in \citet{Kurayama:2011} arise from this weak foreground cloud. The main cloud is quite large and well studied; the masers lie in known regions of high-mass star formation which lie precisely along the dark filament we identify for use in the extinction distance. The near-infrared data does not appear abnormal in any way. 

This source remains a puzzle. It is possible that the maser parallax distance is incorrect, or at least that the errors are significantly underestimated. Because this source is very far south for the VERA array (+1 \arcdeg\ of declination), the declination does not constrain the parallax, which comes only from fitting right ascension \citep{Kurayama:2011}. In addition, the model for the right ascension data (annual parallax plus a linear proper motion) is a poor fit to the observations, with 8 out of 21 observations excluded from the fit for the three maser features used. Additional maser parallax measurements of this source would be highly desirable.

\section{Summary}

New extinction distance measurements have been made toward 11 dark molecular clouds for which direct distances are available from maser parallax measurements. The two extinction methods are the Blue Number Count method and the Red Giant method. In addition, we have determined kinematic distances to these clouds using dense gas tracers and two different rotation curves \citep[][]{Clemens:1985, Reid:2009}. 

The Blue Number Count method presented here uses auxiliary information to identify the boundaries of a cloud within which there is enough column density to cleanly separate all foreground stars from background stars. The number of blue foreground stars within these boundaries is then compared to a Galactic model of the stellar distribution to estimate the distance to the cloud. Because the surface density of blue stars must be estimated accurately we have used injection and recovery of synthetic stars to estimate the completeness as a function of magnitude for each region (for both 2MASS and UKIDSS). 

The Blue Number Count method requires some additional information in order to determine if there are multiple clouds along the line of sight; if there are multiple clouds along the line of sight the method can only estimate a distance to the nearest cloud. The Blue Number Count method also only works for clouds with sufficiently high column density over a large enough region to have a significant number of blue foreground stars in whichever near-infrared catalog is being used. Possible sources of systematic error include multiple clouds along the line of sight, inhomogeneities in the stellar distribution, errors introduced by the completeness estimation and photometry routines, and deficiencies in the Galactic model used.

Our reliance on stellar number density means that the Blue Number Count method tests a different aspect of the Besan\c con model than \citet{Marshall:2009}, which uses only color information and discards number density information by normalizing within each color bin. \citet{Marshall:2009} found a significant (1.5 kpc) systematic offset between extinction distances and kinematic distances in the fourth quadrant. Although this work does not directly address this offset, since none of the clouds with maser parallax distance determinations are in the fourth quadrant, the use of complimentary information from the Besan\c con model will allow us to examine the \citet{Marshall:2009} result in the fourth quadrant in future work.

The Red Giant method used here is similar to that presented in \citet{stead10}, whereby individual extinction measurements are made to color-selected giant stars along the same line of sight as molecular clouds. Following this, extinction histograms are constructed and molecular clouds can be identified as large gaps in what would otherwise be a relatively smooth distribution. Using an existing extinction-distance relation (EDR) the extinction measurements can be converted to reliable distances. However this method suffers at near distances where an inadequate number of giant stars in front of the cloud are available. The EDRs used here are those of \citet{Marshall:2006}. As they gridded the Galaxy into 15$^{\prime}$x15$^{\prime}$ tiles to derive their EDRs, the spatially closest tile to each cloud has been used. The reliability of this method could therefore be improved if line of sight specific EDRs were derived using the very same stars used to build the extinction histogram. 

In some cases multiple molecular clouds have been identified along the same line of sight; such cases pose a challenge for extinction distance determinations. In this sample, multiple molecular clouds are typically observed towards the Galactic center where the surface density of clouds is high. The Blue Number Count method can only determine the distance to the first cloud along a line of sight, and may fail even for the foreground cloud if the multiple clouds along the line of sight cause us to misidentify the region which is dense enough to produce a clear separation between background and foreground stars. The Red Giant method can, with sufficient number of stars, identify multiple clouds along the line of sight, a clear advantage for this method in crowded regions. 

Our individual distance determinations are presented in Table~\ref{Table:Distances}. The extinction distance methods agree to within 2$\sigma$ with the maser parallax distances between 66\% and 100\% of the time (depending on method and input survey), and between 85\% and 100\% of the time outside of the crowded Galactic center (Table~\ref{comparison}). Extinction distance estimates provide better agreement with maser parallax distances than the kinematic distances obtained using either of two different rotation curves. Extinction distance methods provide an important, independent, estimate of the distance to dark molecular clouds. These methods do fail occasionally, but normally there is a good explanation such as multiple clouds along the line of sight, which can be established through auxiliary information. In a small number of cases it is possible that that maser parallax distance is incorrect, or is reported with error bars which are too small. In the absence of a maser parallax distance, the combination of both extinction and kinematic distances will provide the most secure measurement of the distance to a cloud.

\section{Acknowledgements}
J.B.F. gratefully acknowledges support from NASA grant NNX09AC82G. R.A.B. acknowledges support from NASA grant NNX10AI70G . This publication makes use of data products from the Two Micron All Sky Survey, which is a joint project of the University of Massachusetts and the Infrared Processing and Analysis Center/California Institute of Technology, funded by the National Aeronautics and Space Administration and the National Science Foundation. This work is based in part on data obtained as part of the UKIRT Infrared Deep Sky Survey. This research has made use of NASA's Astrophysics Data System. Thanks to useful discussions with Mark Reid and Patricio Sanhueza.

\clearpage
\appendix
\label{sec:Appendix}
\section{Blue Number Count Extinction Distance Figures}
\clearpage

\begin{figure}
  \begin{center}
    \begin{tabular}{ccc}
      \resizebox{57mm}{!}{\includegraphics[angle=0]{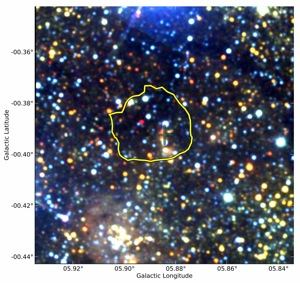}}&
     \resizebox{57mm}{!}{\includegraphics[angle=0]{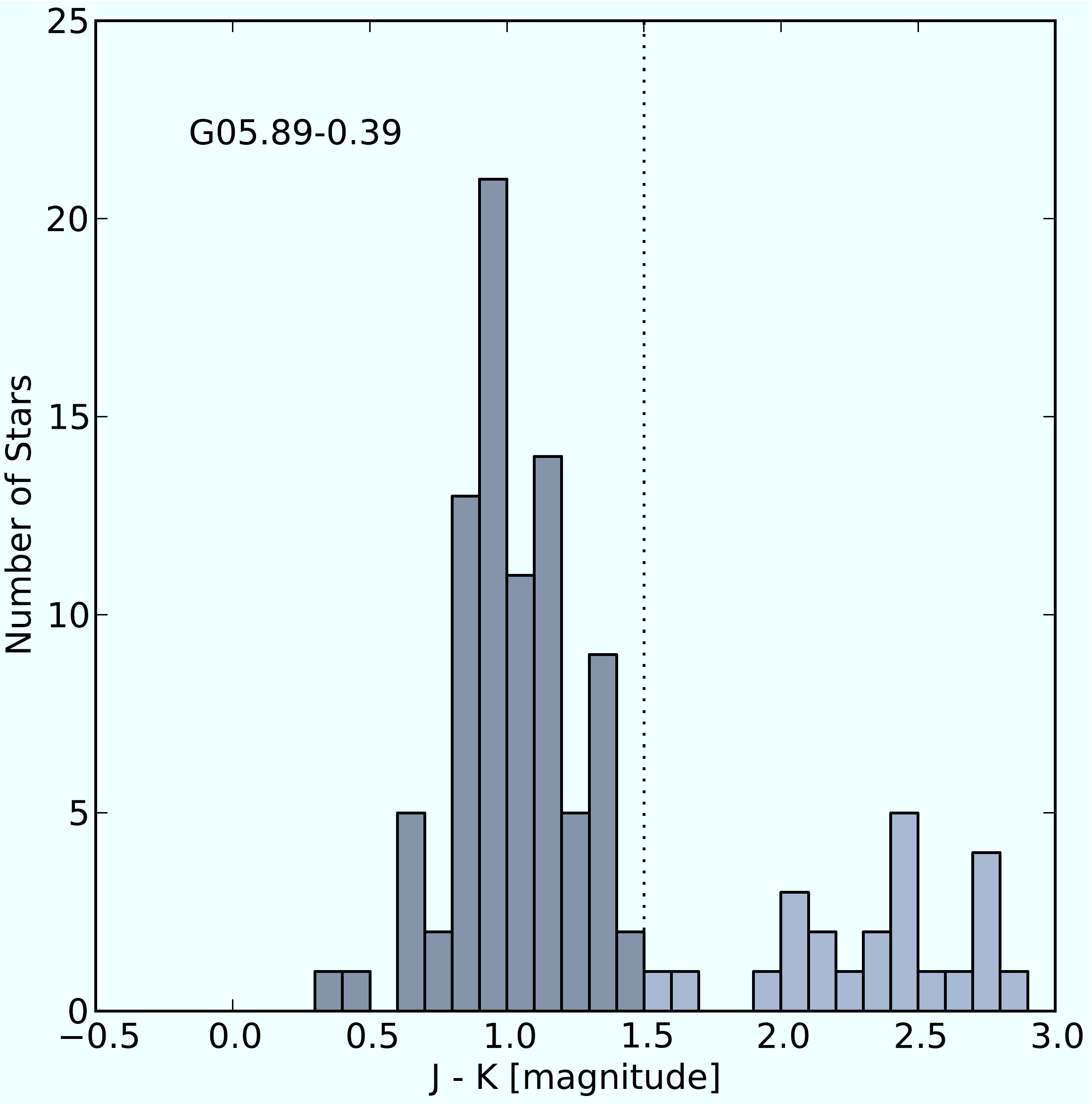}}&
     \resizebox{57mm}{!}{\includegraphics[angle=0]{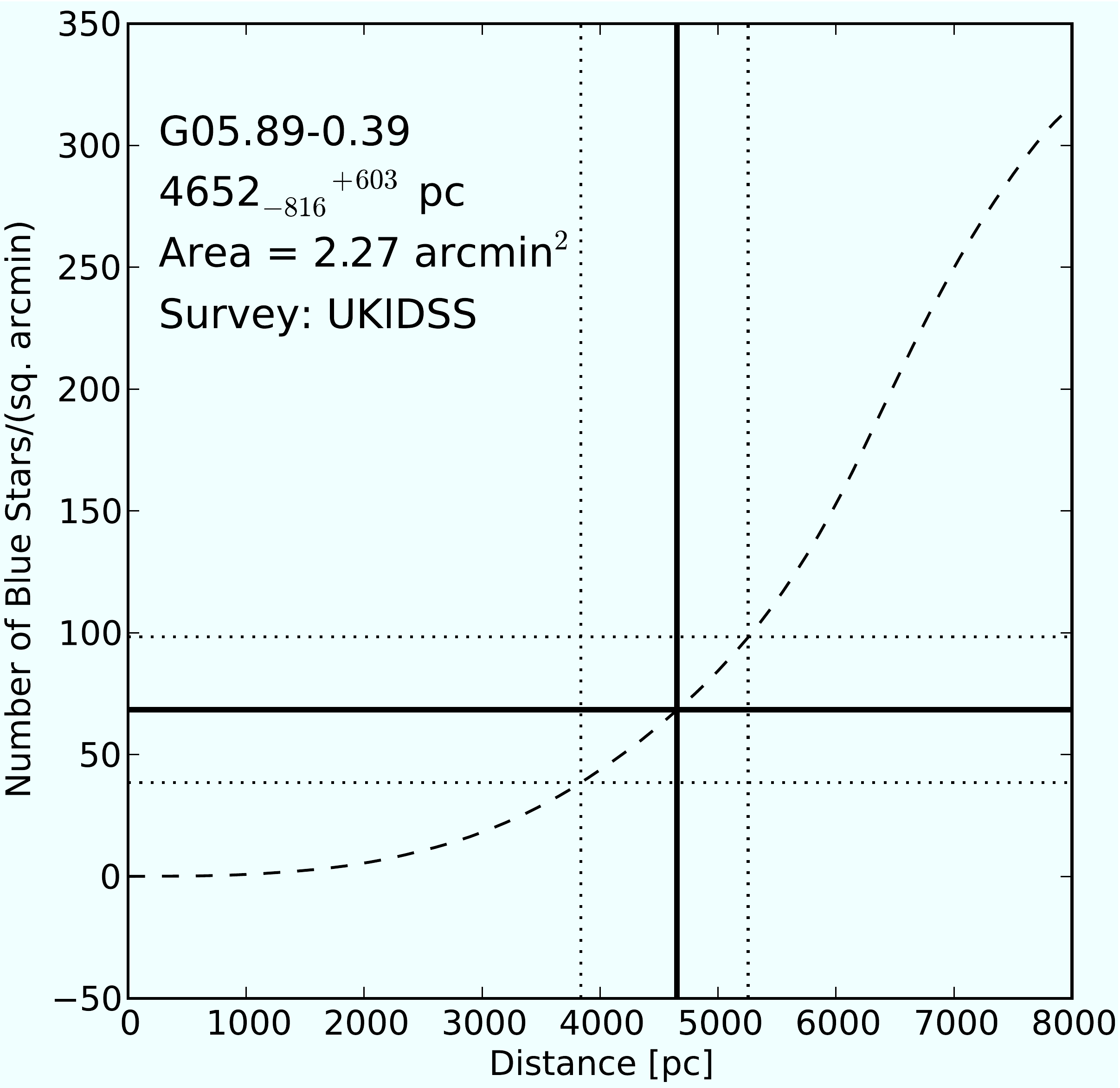}}\\
     \resizebox{57mm}{!}{\includegraphics[angle=0]{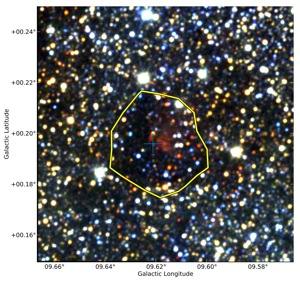}}&
     \resizebox{57mm}{!}{\includegraphics[angle=0]{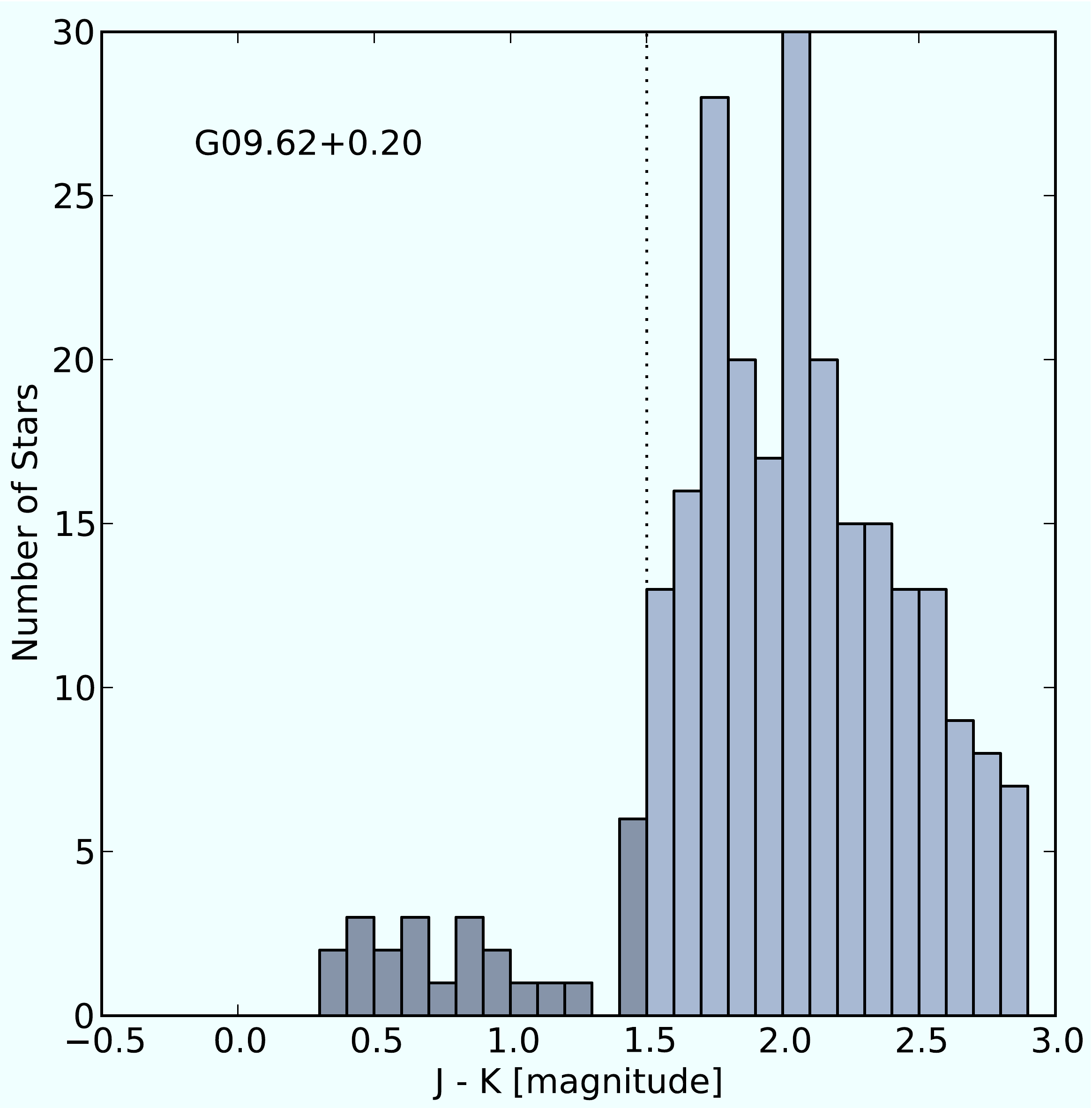}}&
     \resizebox{57mm}{!}{\includegraphics[angle=0]{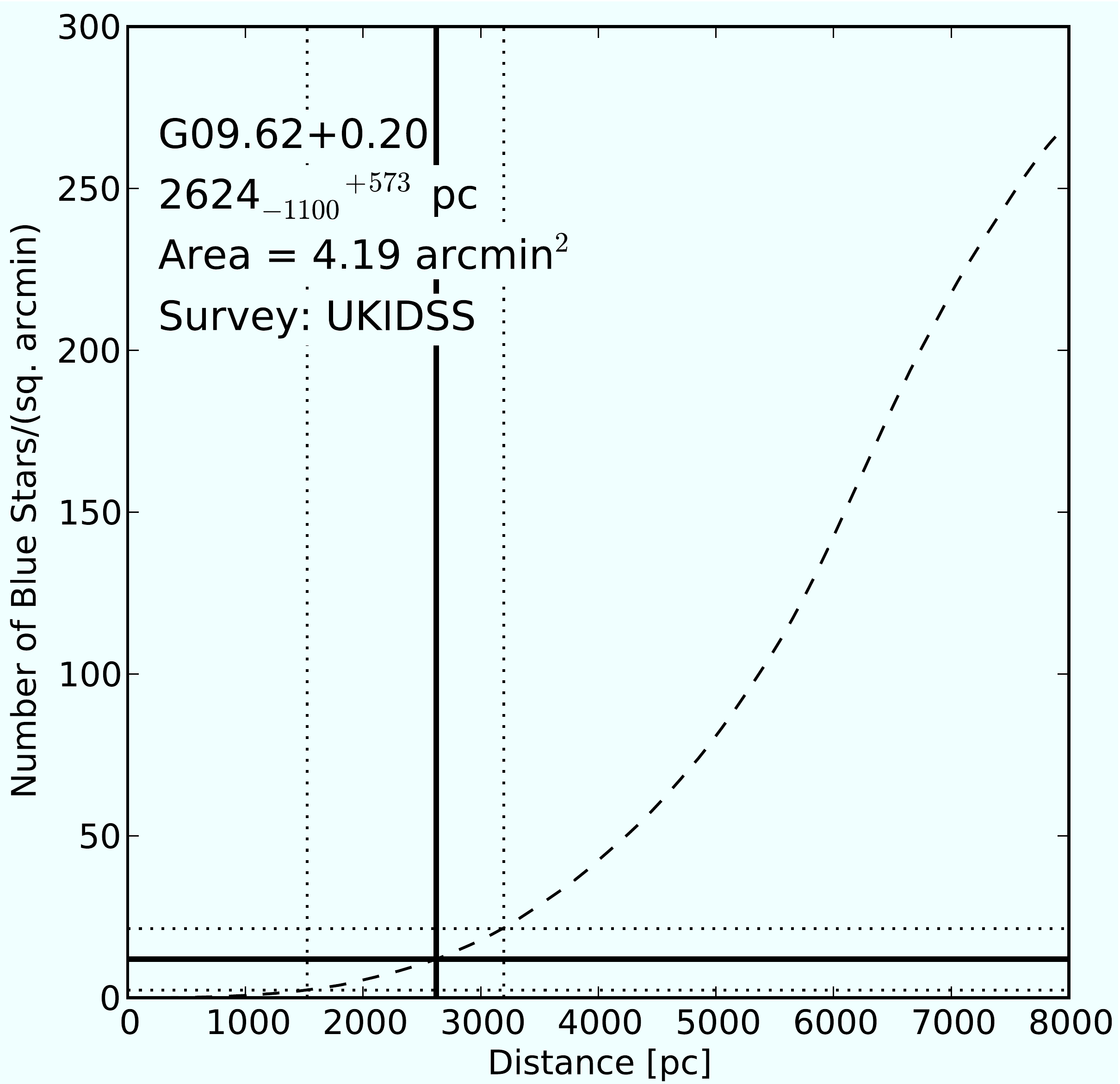}} \\
     \resizebox{57mm}{!}{\includegraphics[angle=0]{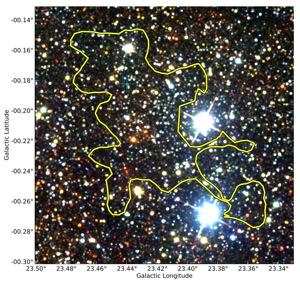}}&
     \resizebox{57mm}{!}{\includegraphics[angle=0]{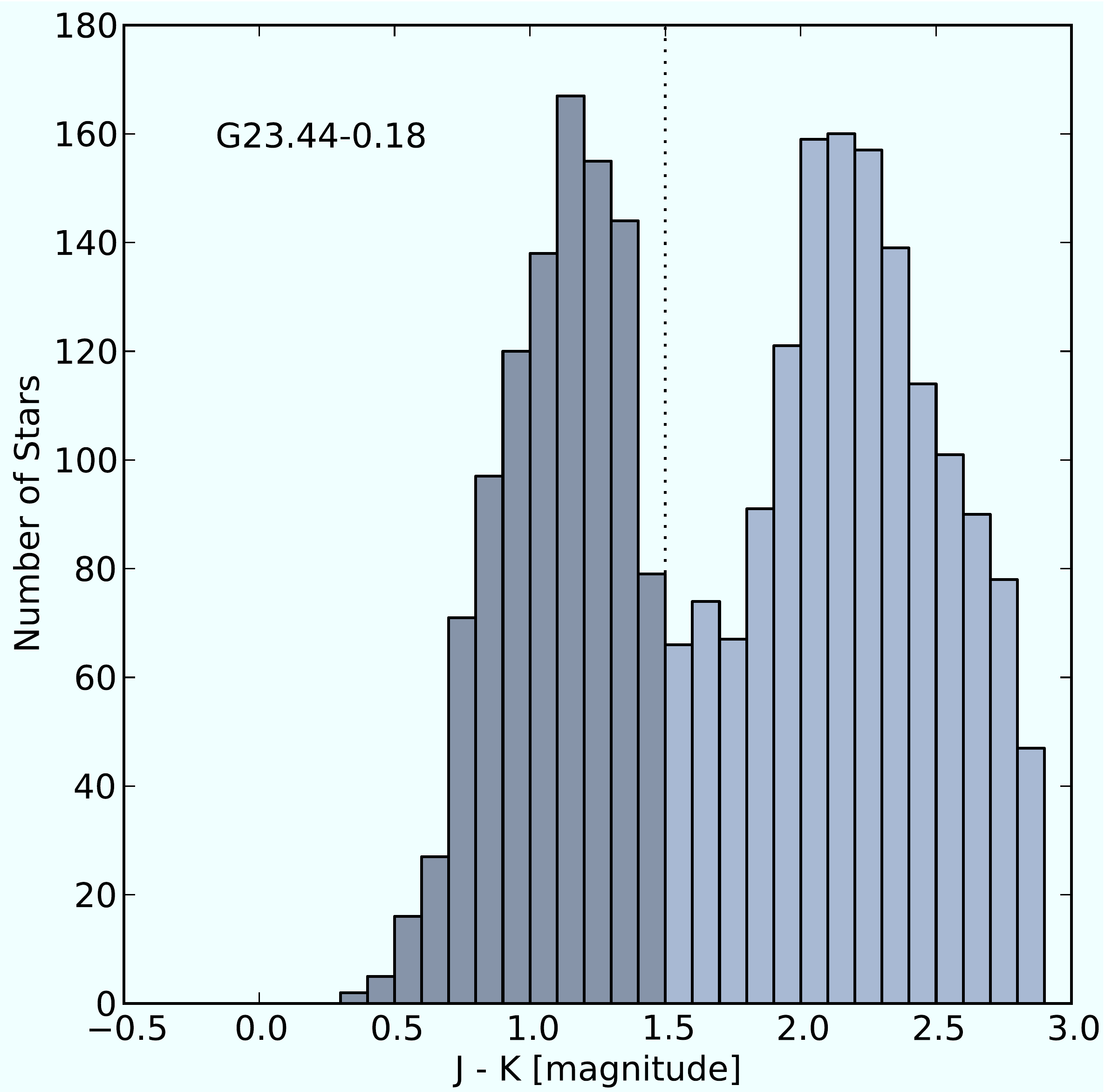}}&
     \resizebox{57mm}{!}{\includegraphics[angle=0]{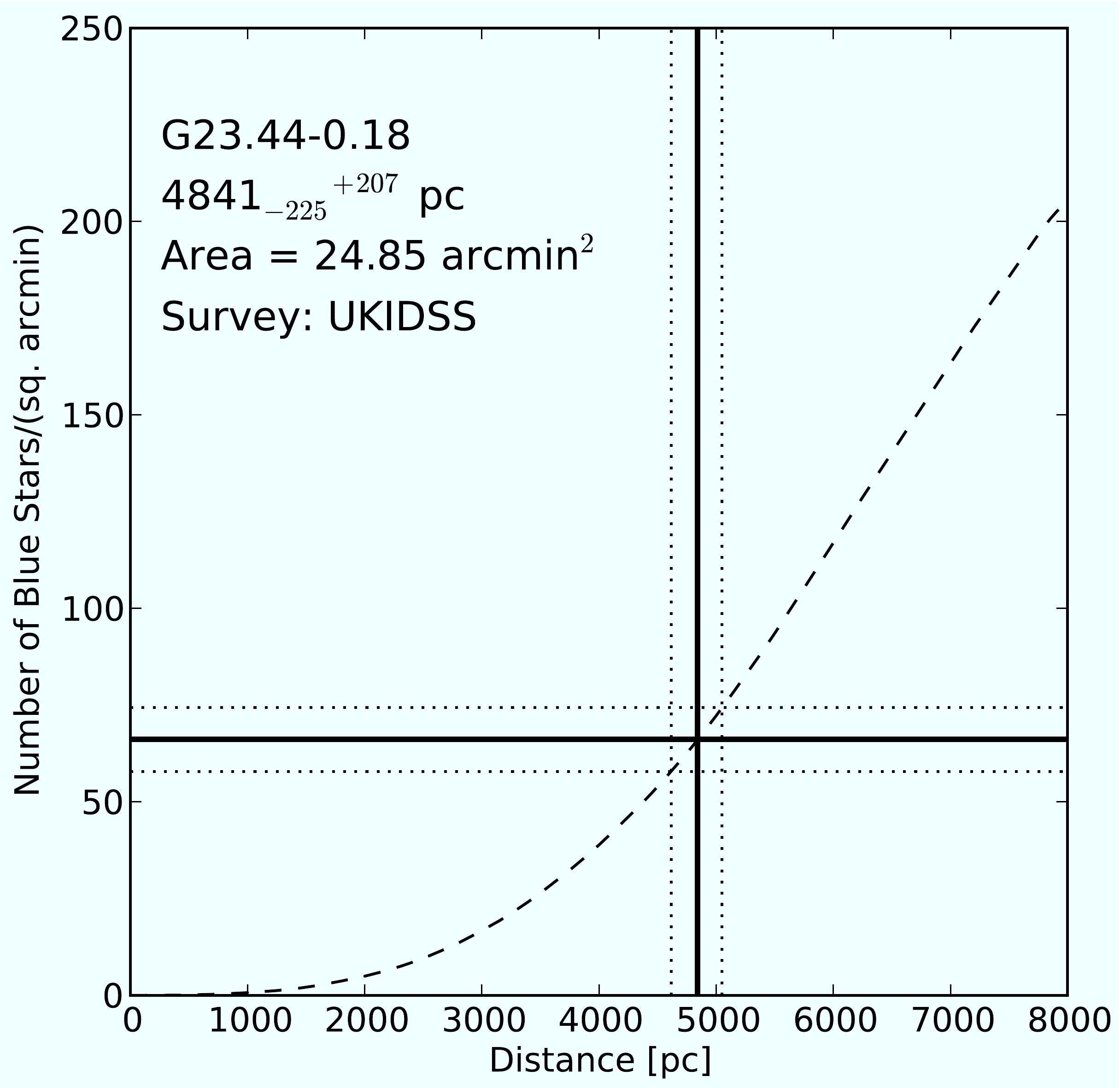}} 
    \end{tabular}
    \caption{Similar to Figure~\ref{SampleImage}, \ref{SampleHist}, and \ref{SampleDetermination}. For each source, we show (left) the 2MASS three-color image ($J$=blue, $H$=green, $K$=red) with the cloud contour and the position of the maser overlain. In the center we show the histogram of UKIDSS (where available, otherwise 2MASS) $J$-$K$ colors inside the cloud boundaries, and show the cutout at $J$-$K$ = 1.5. On the right we show the UKIDSS (where available, otherwise 2MASS) completeness-corrected blue stellar density and how this compares to the Galactic model.}
    \label{fig:G1}
  \end{center}
\end{figure}

\begin{figure}
  \begin{center}
    \begin{tabular}{ccc}
     \resizebox{57mm}{!}{\includegraphics[angle=0]{G5_3color}}&
     \resizebox{57mm}{!}{\includegraphics[angle=0]{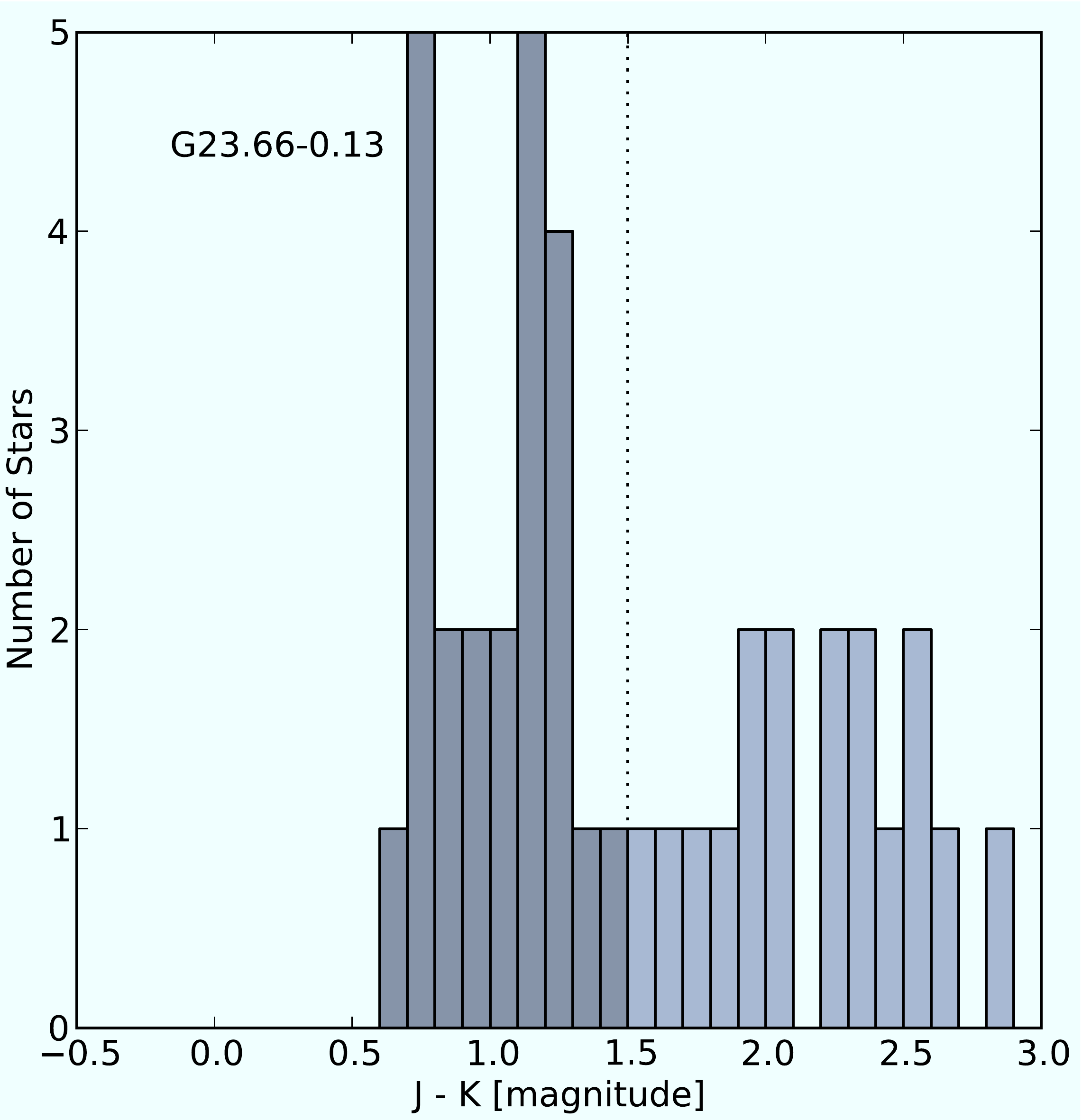}}&
     \resizebox{57mm}{!}{\includegraphics[angle=0]{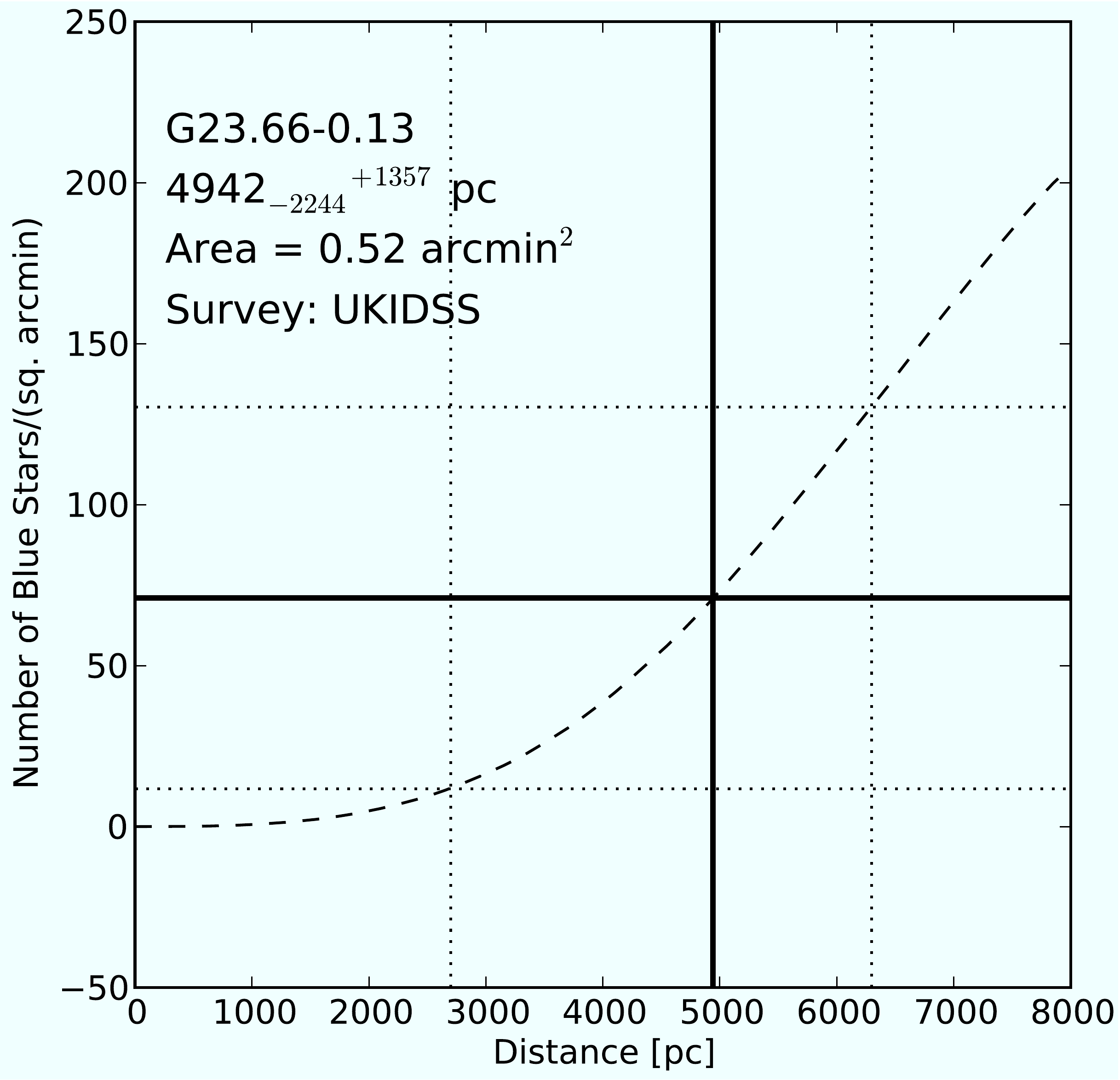}} \\
     \resizebox{57mm}{!}{\includegraphics[angle=0]{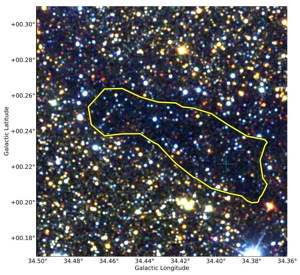}}&
     \resizebox{57mm}{!}{\includegraphics[angle=0]{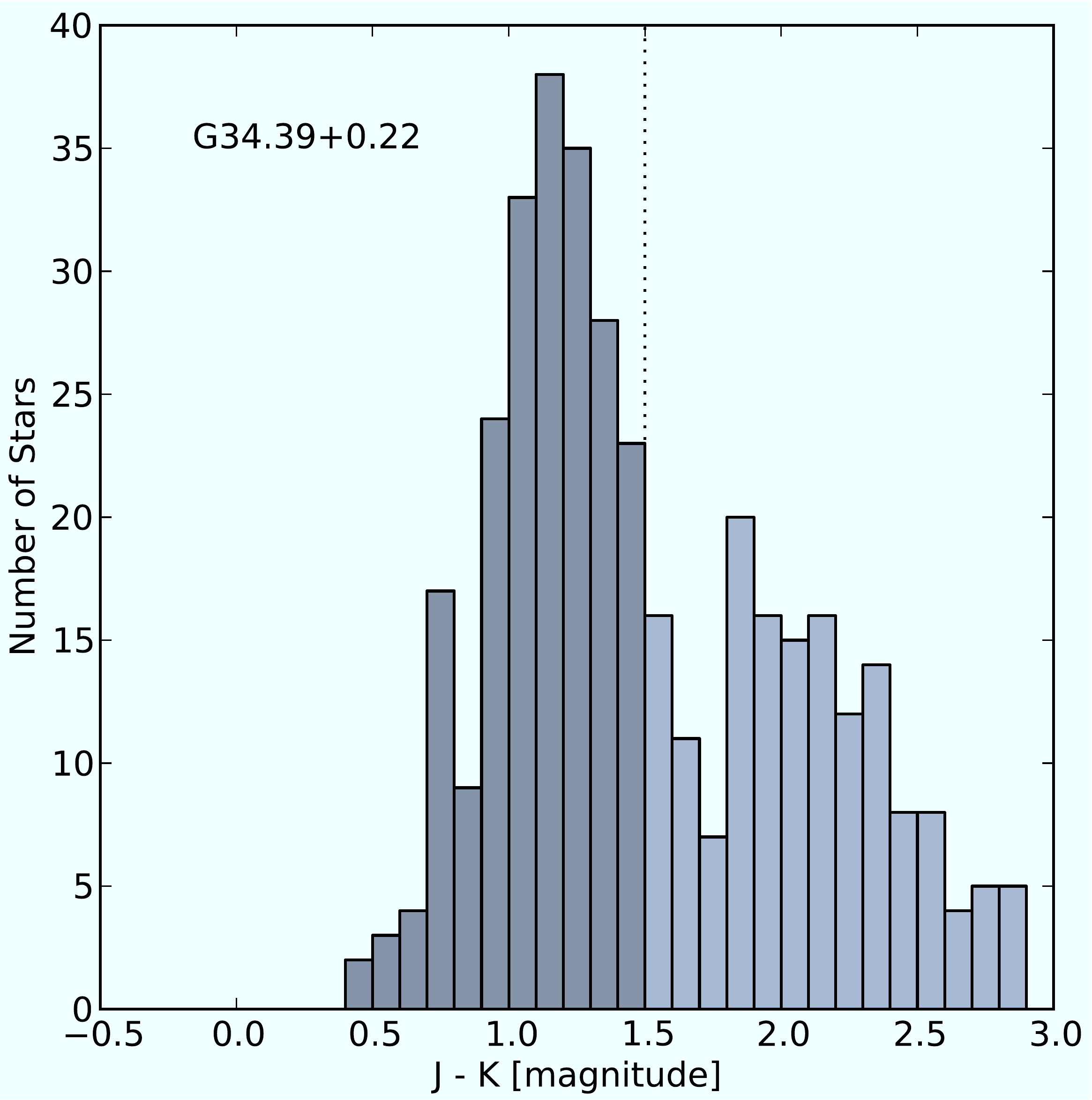}}&
     \resizebox{57mm}{!}{\includegraphics[angle=0]{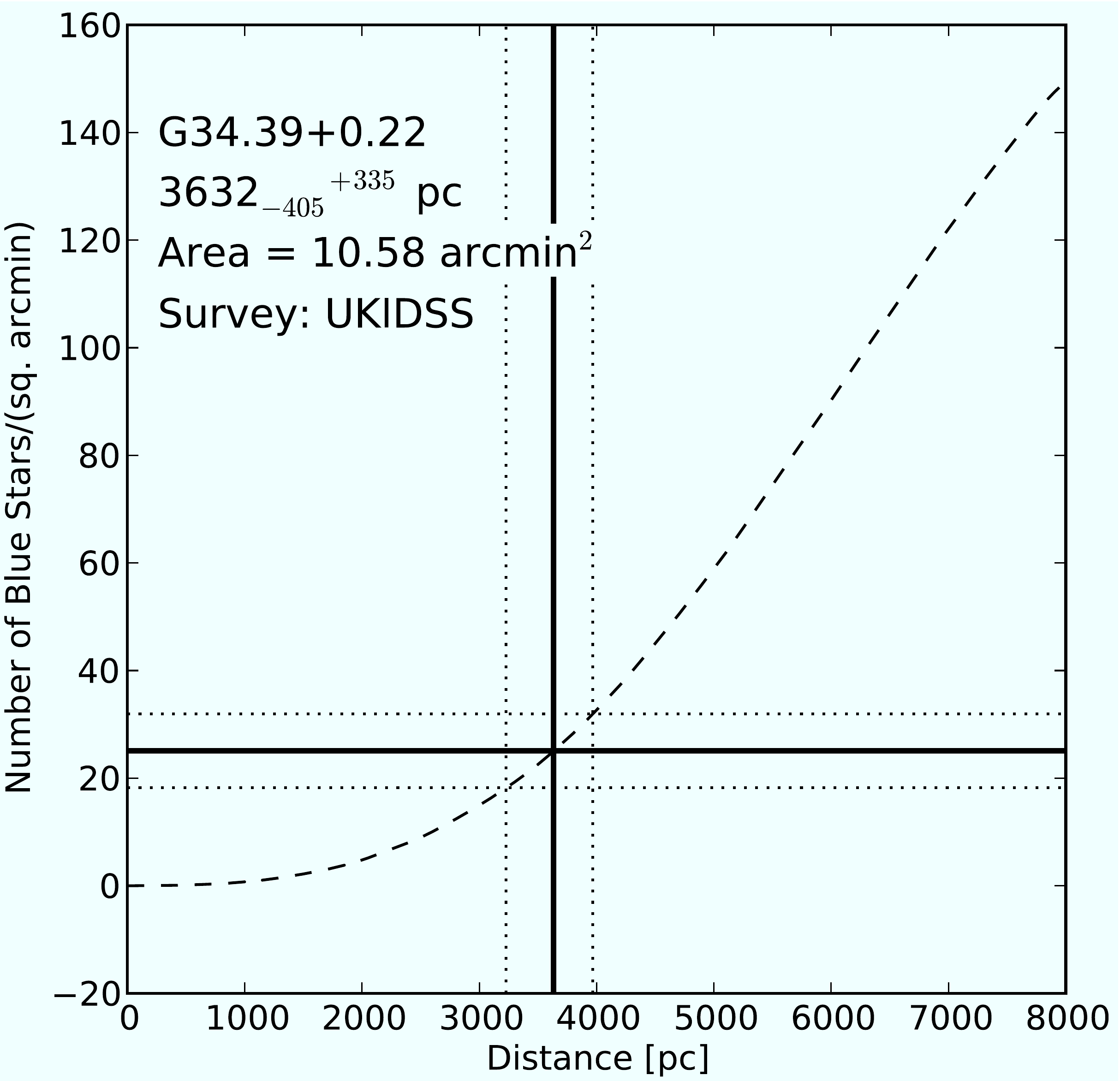}} \\
     \resizebox{57mm}{!}{\includegraphics[angle=0]{G7_3color}}&
     \resizebox{57mm}{!}{\includegraphics[angle=0]{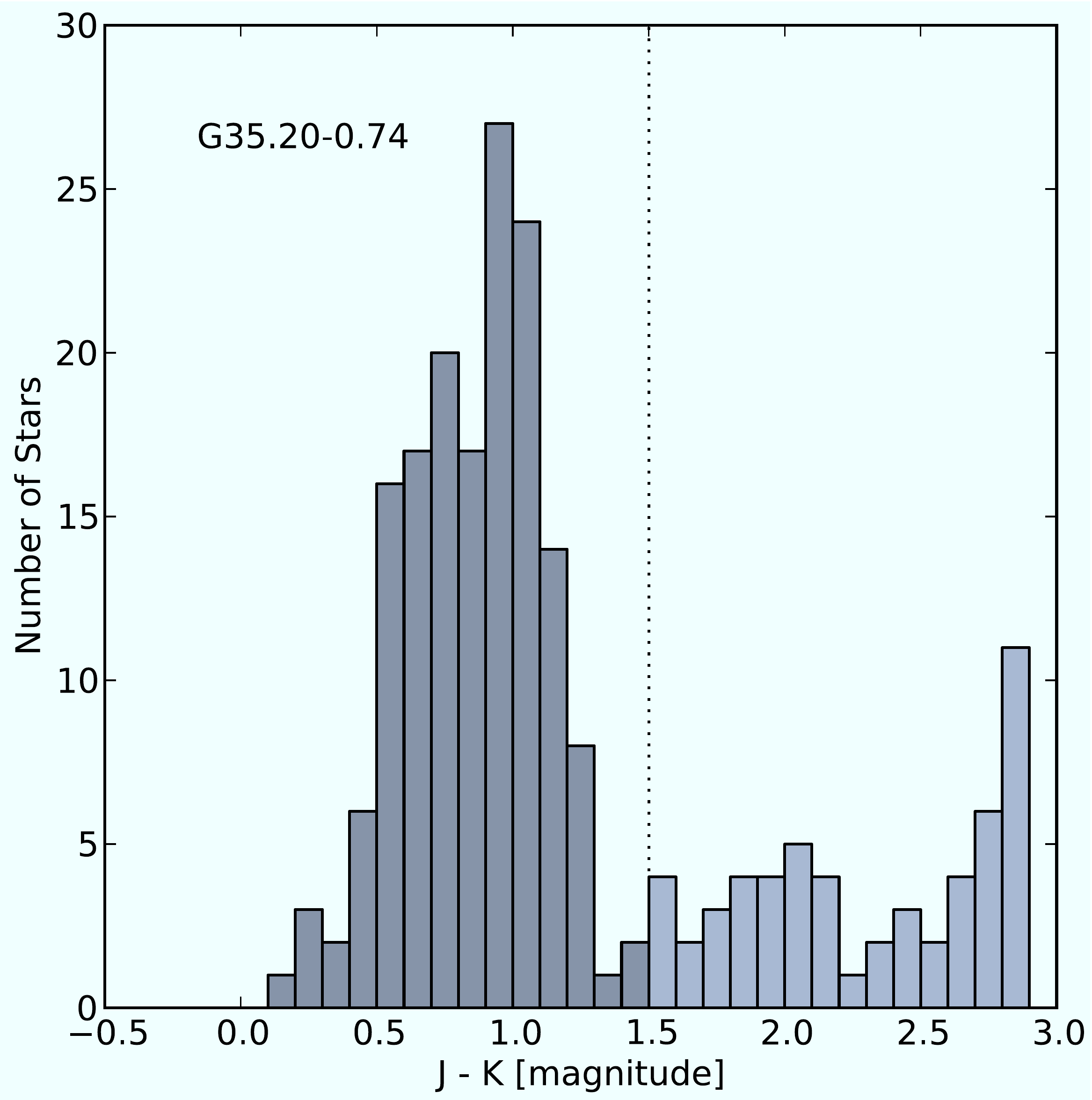}}&
     \resizebox{57mm}{!}{\includegraphics[angle=0]{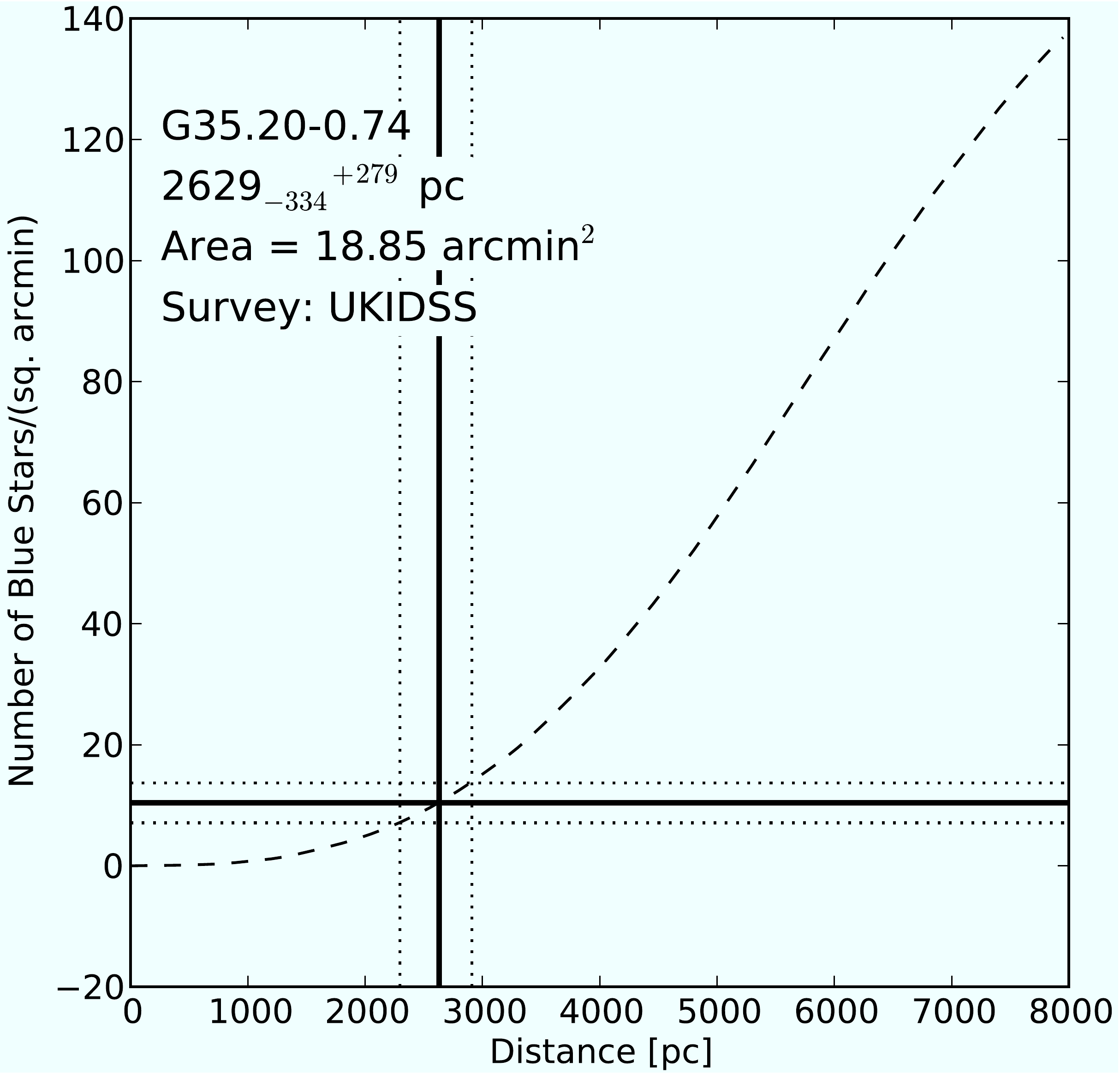}} 
    \end{tabular}
    \caption{Same as Figure~\ref{fig:G1}}
    \label{fig:G2}
  \end{center}
\end{figure}

\begin{figure}
  \begin{center}
    \begin{tabular}{ccc}
     \resizebox{57mm}{!}{\includegraphics[angle=0]{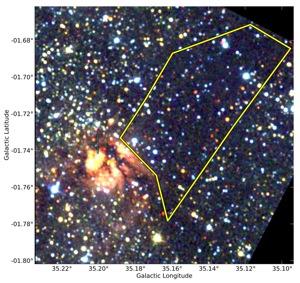}}&
     \resizebox{57mm}{!}{\includegraphics[angle=0]{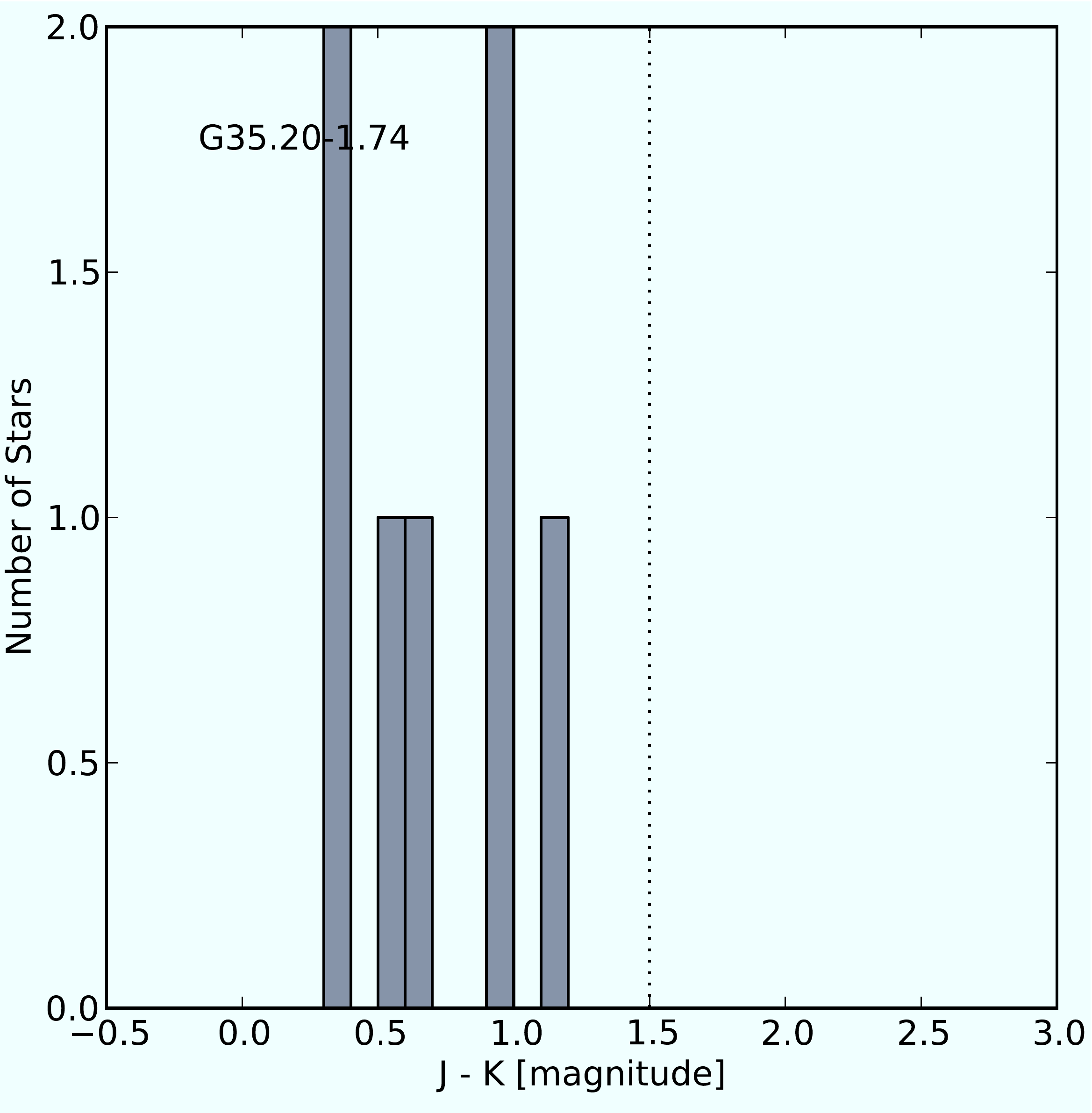}}&
     \resizebox{57mm}{!}{\includegraphics[angle=0]{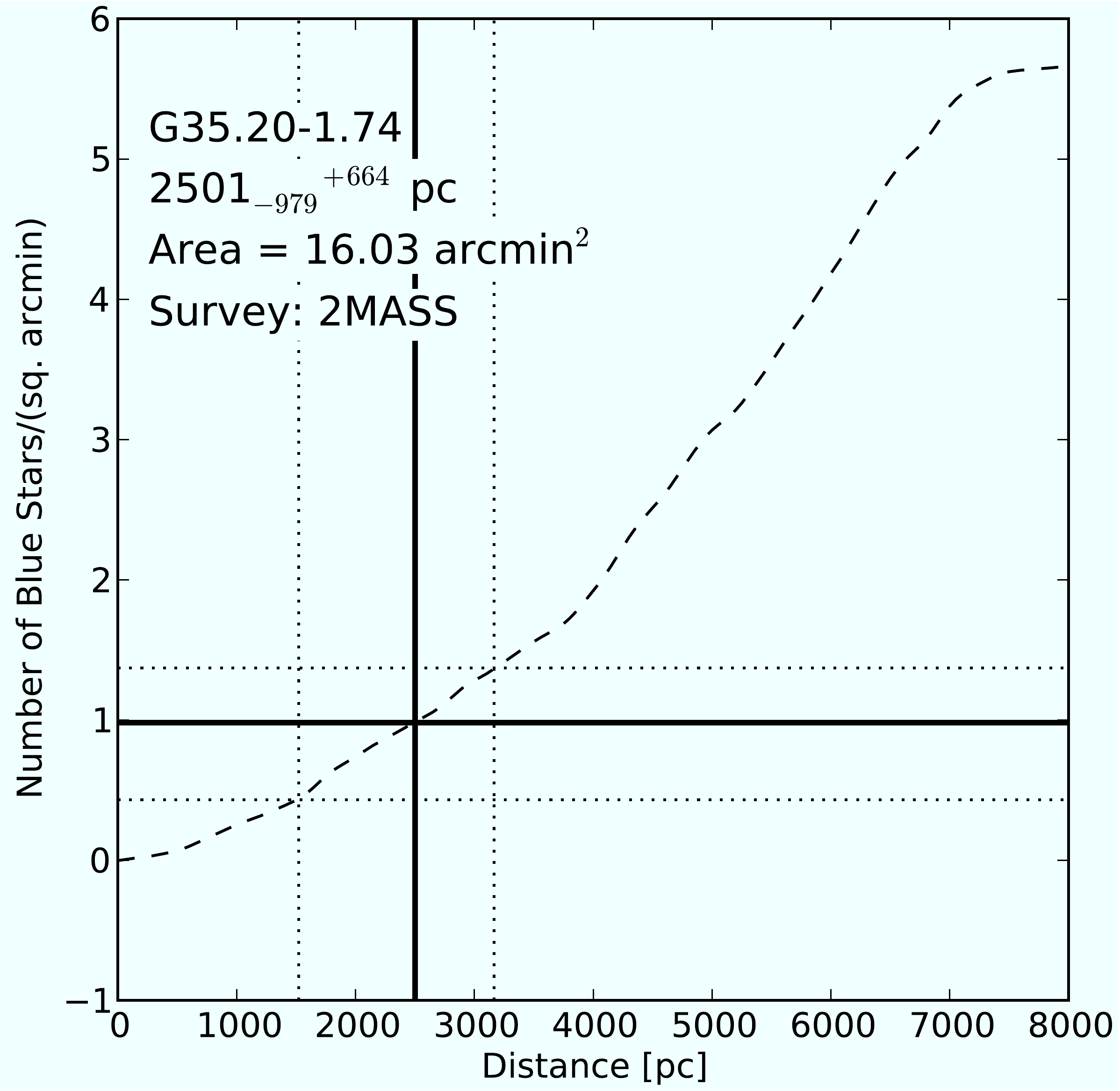}} \\
      \resizebox{57mm}{!}{\includegraphics[angle=0]{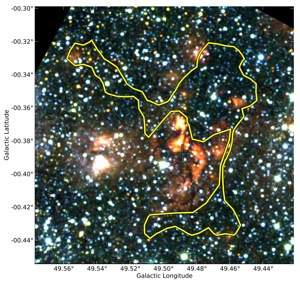}}&
     \resizebox{57mm}{!}{\includegraphics[angle=0]{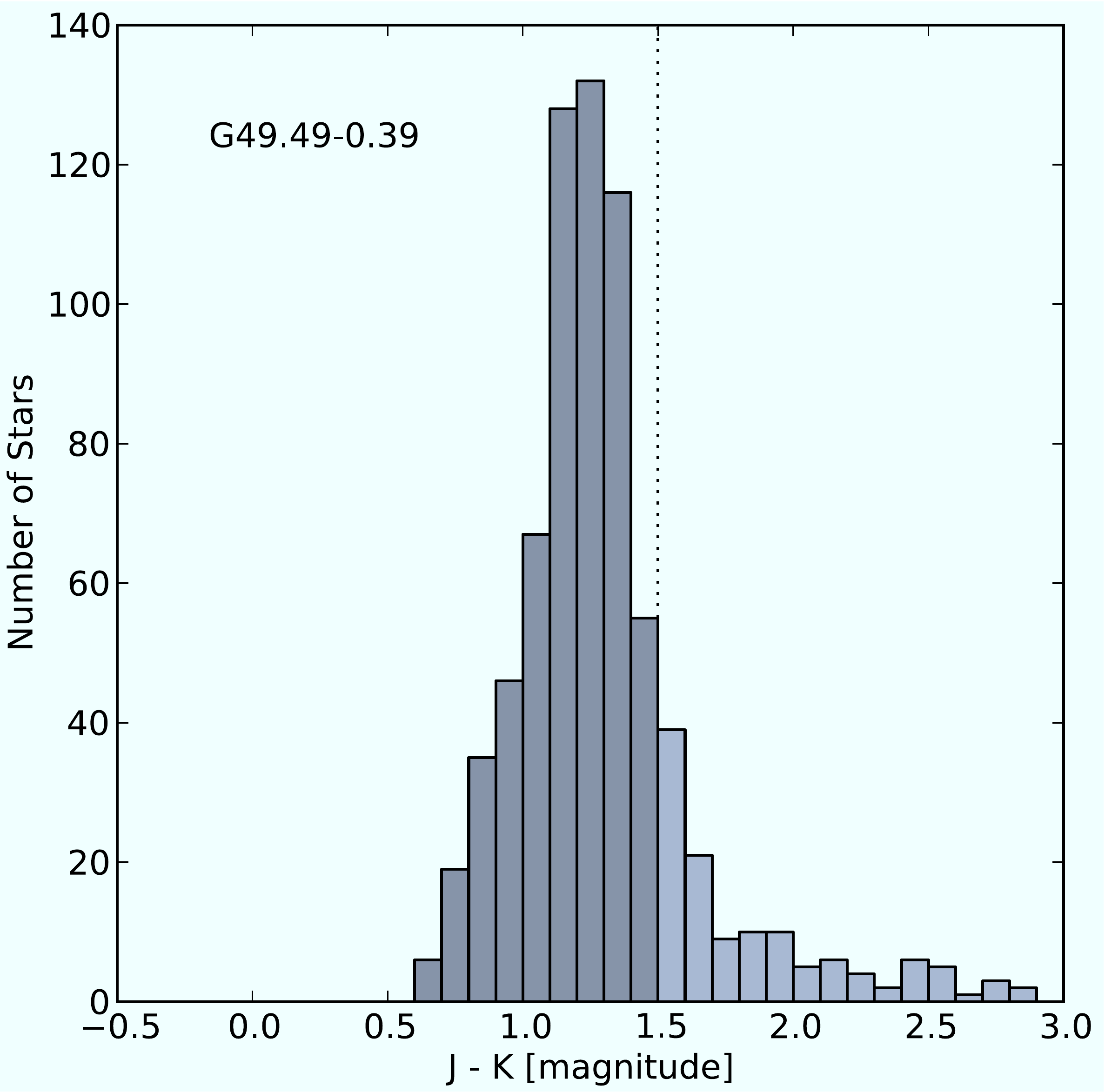}}&
     \resizebox{57mm}{!}{\includegraphics[angle=0]{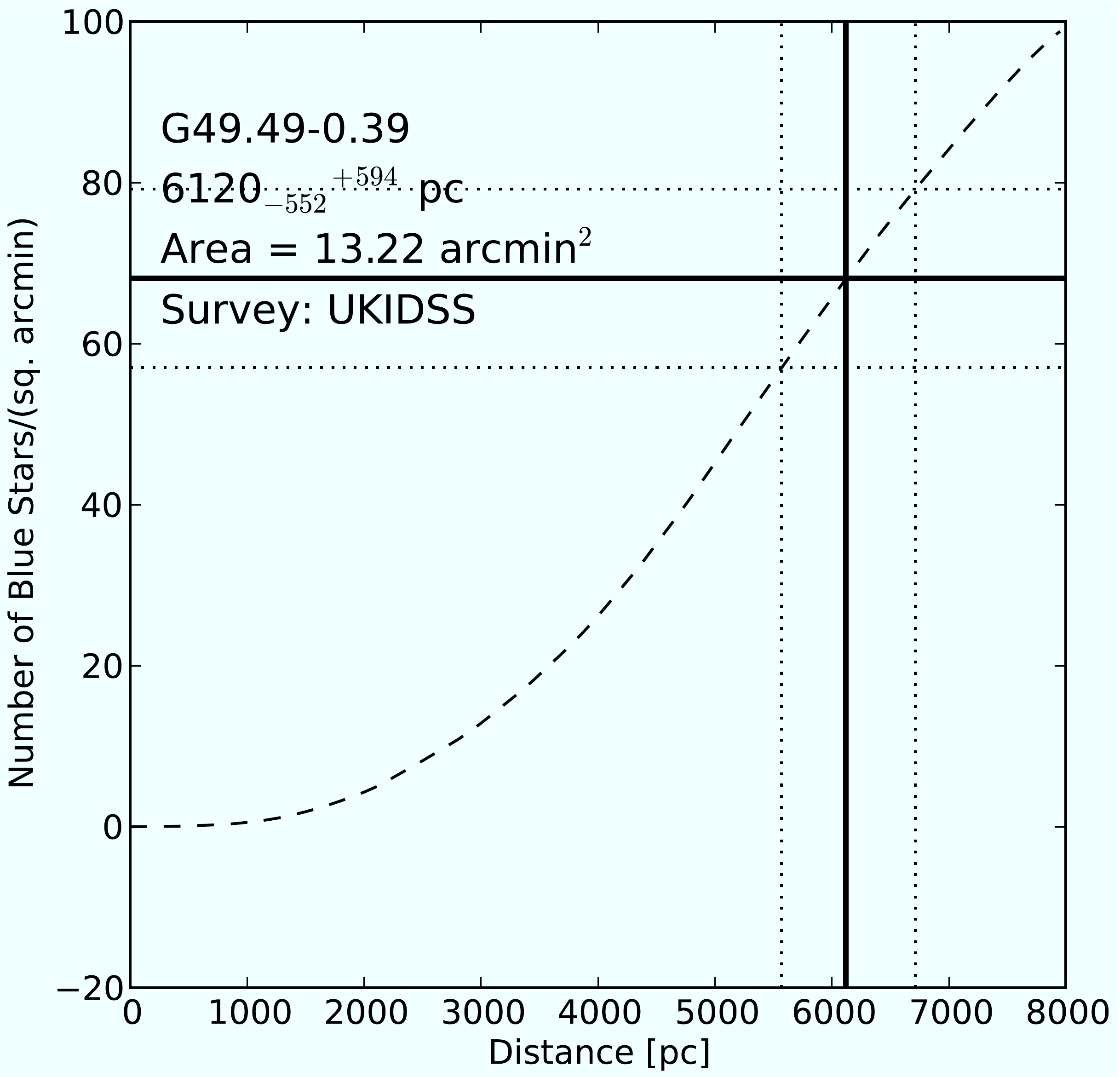}} \\
      \resizebox{57mm}{!}{\includegraphics[angle=0]{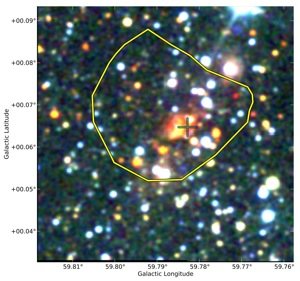}}&
     \resizebox{57mm}{!}{\includegraphics[angle=0]{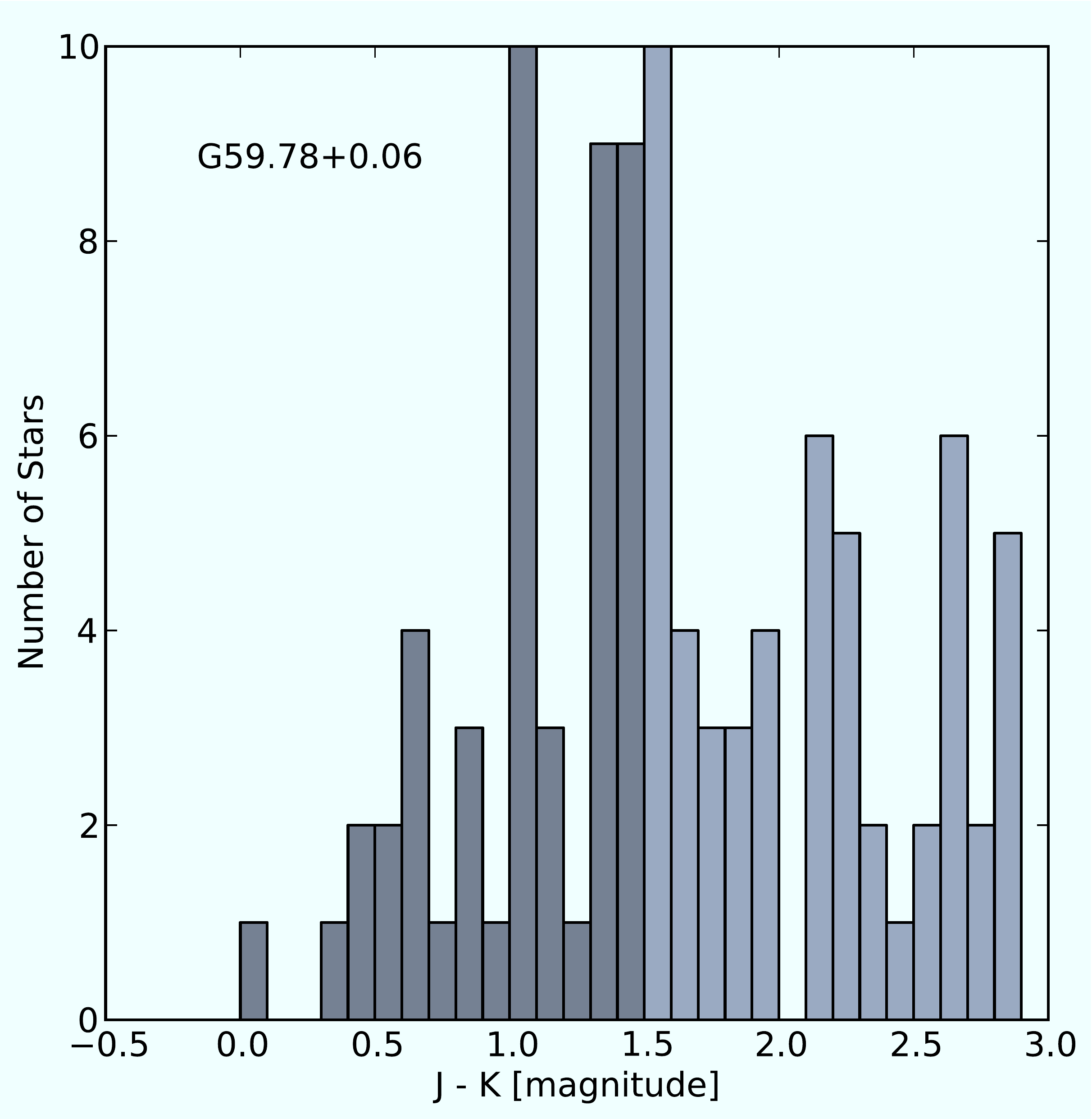}}&
     \resizebox{57mm}{!}{\includegraphics[angle=0]{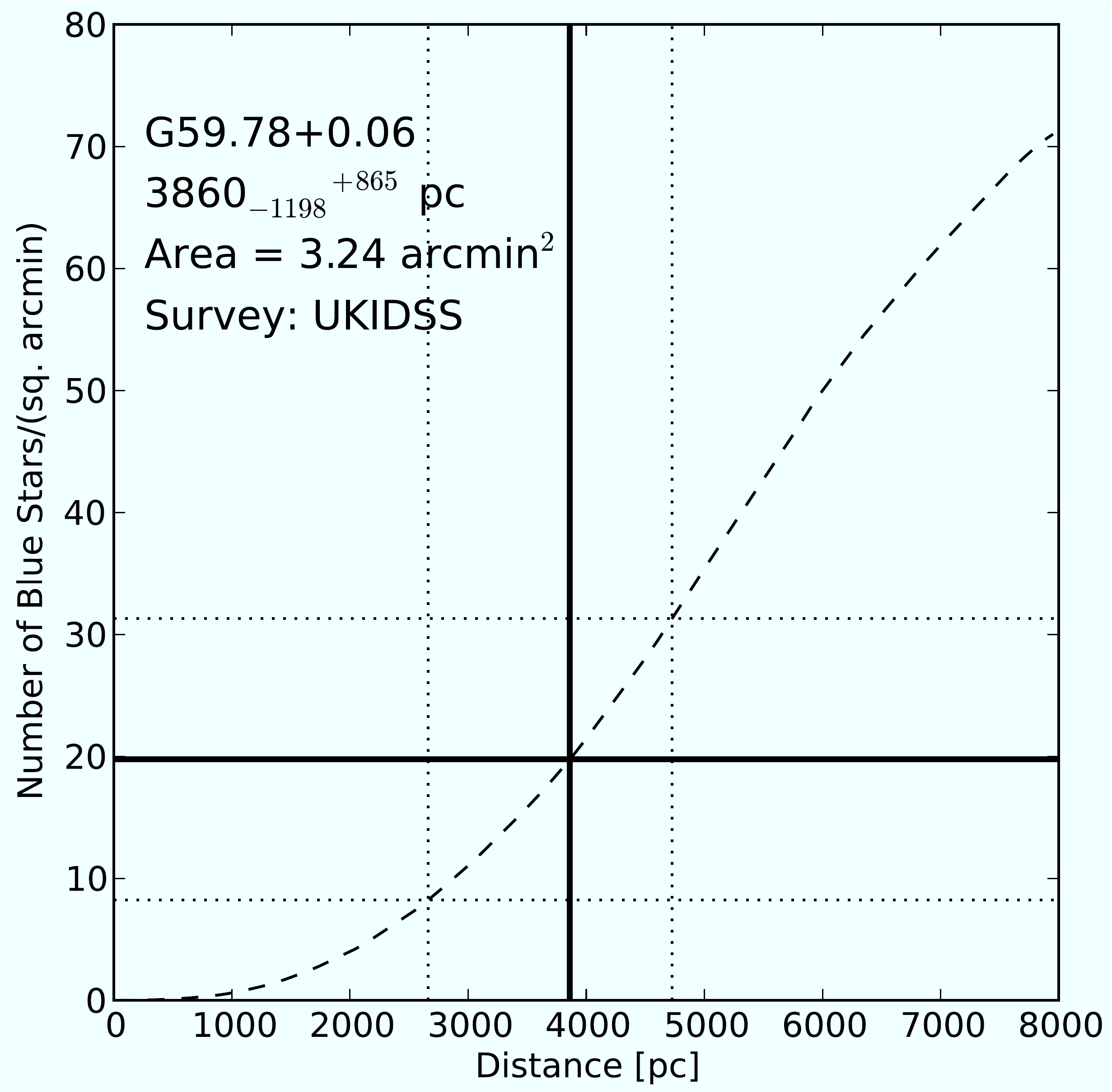}}
    \end{tabular}
    \caption{Same as Figure~\ref{fig:G1}}
    \label{fig:G3}
  \end{center}
\end{figure}

\clearpage
\section{Red Giant Extinction Distance Figures}
\clearpage

\begin{figure*}
  \begin{center}
    \begin{tabular}{ccc}
      \resizebox{57mm}{!}{\includegraphics[angle=0]{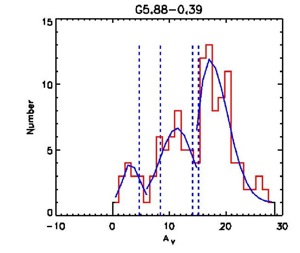}}&
     \resizebox{57mm}{!}{\includegraphics[angle=0]{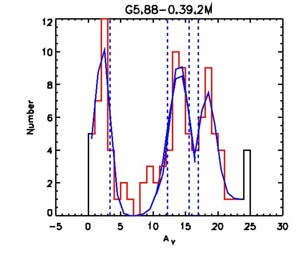}}&
     \resizebox{57mm}{!}{\includegraphics[angle=0]{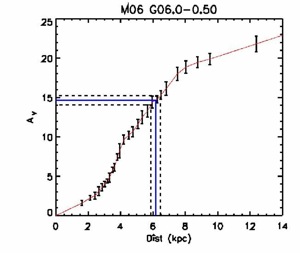}}\\
      \resizebox{57mm}{!}{\includegraphics[angle=0]{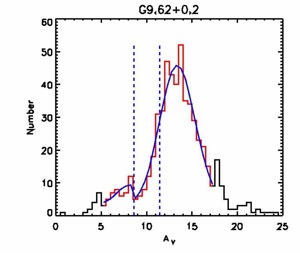}}&
     \resizebox{57mm}{!}{\includegraphics[angle=0]{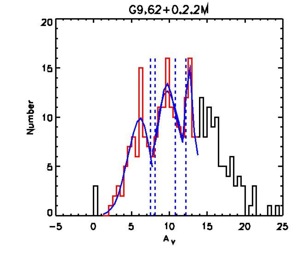}}&
     \resizebox{57mm}{!}{\includegraphics[angle=0]{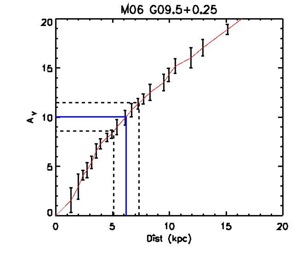}}\\
      \resizebox{57mm}{!}{\includegraphics[angle=0]{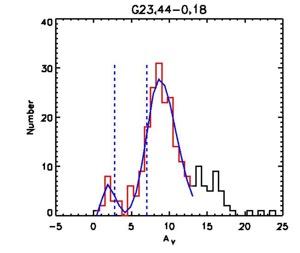}}&
     \resizebox{57mm}{!}{\includegraphics[angle=0]{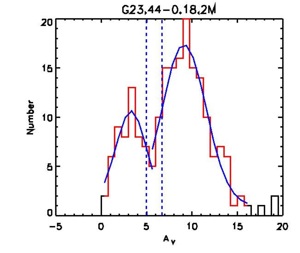}}&
     \resizebox{57mm}{!}{\includegraphics[angle=0]{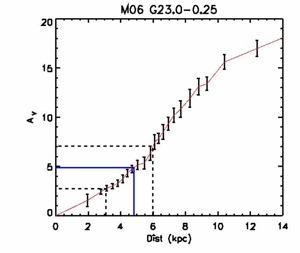}}\\
    \end{tabular}
    \caption{\small Similar to Figure \ref{fig:G23.0096-0.4105_Avhist} (a) and (b), each A$_V$ histogram has been labeled with the name of the target cloud - the 2MASS histograms have been denoted with a `2M'. The near and far sides of each molecular cloud are marked with dashed lines, derived from the 1$\sigma$ deviations of the skewed Gaussians fitted to each histogram (solid curve). Distance-A$_V$ plots have also been created, similar to Fig. \ref{fig:G23.0096-0.4105_Avhist} (c), showing the spatially closest EDR to each cloud (solid thin line). The two sets of dashed lines illustrate the near and far sides of a molecular cloud.}
    \label{fig:A1}
  \end{center}
\end{figure*}
\begin{figure*}
  \begin{center}
    \begin{tabular}{ccc}
          \resizebox{57mm}{!}{\includegraphics[angle=0]{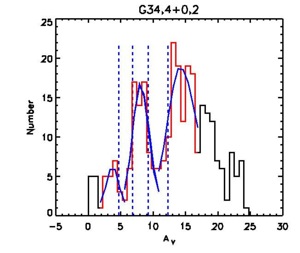}}&
     \resizebox{57mm}{!}{\includegraphics[angle=0]{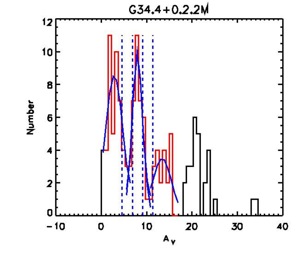}}&
     \resizebox{57mm}{!}{\includegraphics[angle=0]{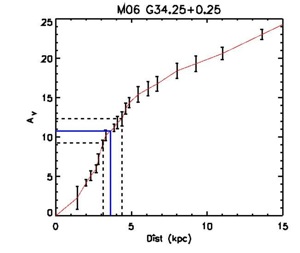}}\\
      \resizebox{57mm}{!}{\includegraphics[angle=0]{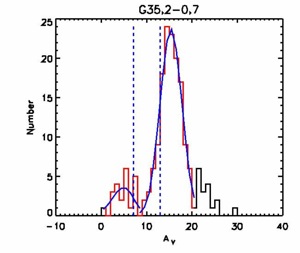}}&
     \resizebox{57mm}{!}{\includegraphics[angle=0]{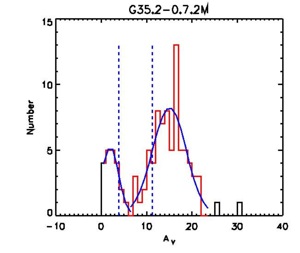}}&
     \resizebox{57mm}{!}{\includegraphics[angle=0]{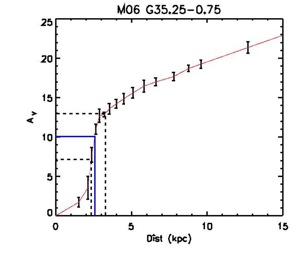}}\\
      \resizebox{57mm}{!}{\includegraphics[angle=0]{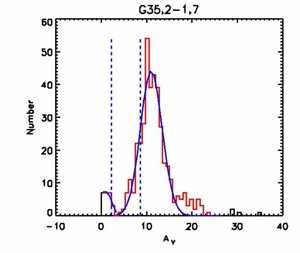}}&
     \resizebox{57mm}{!}{\includegraphics[angle=0]{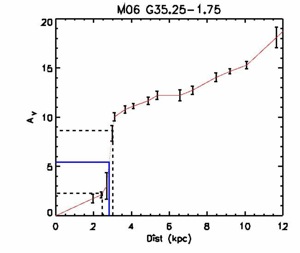}}
    \end{tabular}
    \caption{\small Same as Fig. \ref{fig:A1}.}
    \label{fig:A2}
  \end{center}
\end{figure*}
\begin{figure*}
  \begin{center}
    \begin{tabular}{ccc}
          \resizebox{57mm}{!}{\includegraphics[angle=0]{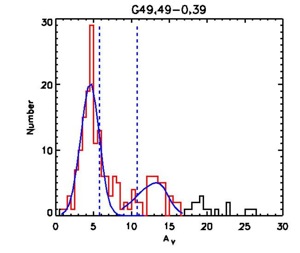}}&
     \resizebox{57mm}{!}{\includegraphics[angle=0]{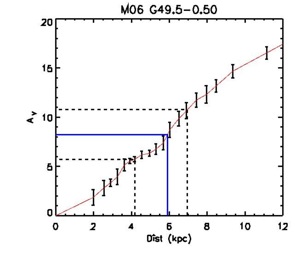}}\\
      \resizebox{57mm}{!}{\includegraphics[angle=0]{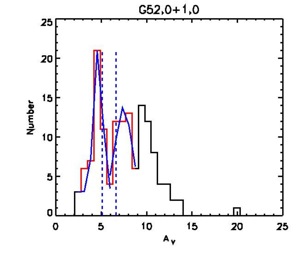}}&
     \resizebox{57mm}{!}{\includegraphics[angle=0]{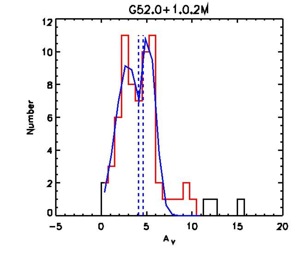}}&
     \resizebox{57mm}{!}{\includegraphics[angle=0]{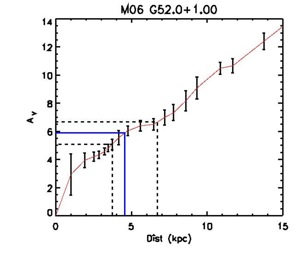}}\\
      \resizebox{57mm}{!}{\includegraphics[angle=0]{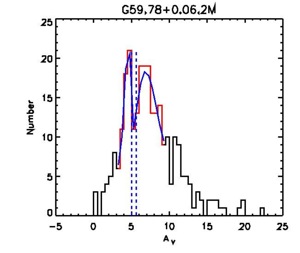}}&
     \resizebox{57mm}{!}{\includegraphics[angle=0]{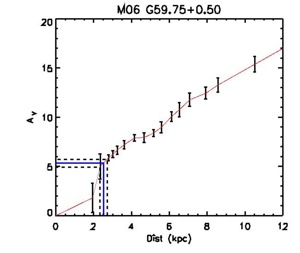}}
    \end{tabular}
    \caption{\small Same as Fig. \ref{fig:A1}.}
    \label{fig:A3}
  \end{center}
\end{figure*}

\end{document}